\documentclass[%
 aip,
 amsmath,amssymb,
preprint,%
]{revtex4-1}

\usepackage{graphicx}
\usepackage{dcolumn}
\usepackage{bm}

\usepackage[utf8]{inputenc}
\usepackage[T1]{fontenc}
\usepackage{mathptmx}
\usepackage{etoolbox}
\usepackage{xcolor}
\usepackage{tikz}
\usetikzlibrary{arrows.meta,arrows}
\usepackage{multirow}
\usepackage[caption=false]{subfig}
\usepackage{hyperref}
\hypersetup{hidelinks}
\usepackage[capitalise]{cleveref}
\usepackage[normalem]{ulem} 
\usepackage{CJK} 
\usepackage{xspace}
\usepackage{framed}
\usepackage{todonotes}

\usepackage{glossaries-extra}
\setabbreviationstyle[acronym]{long-short}
\glssetcategoryattribute{acronym}{nohyperfirst}{true}
\newacronym{RANS}{RANS}{Reynolds-averaged Navier-Stokes}
\newacronym{CFD}{CFD}{Computational Fluid Dynamics}
\newacronym{EPF}{EPF}{Eigenspace Perturbation Framework}
\newacronym{AIM}{AIM}{Anisotropy Invariant Map}
\newacronym{DLR}{DLR}{German Aerospace Center}
\newacronym{UQ}{UQ}{Uncertainty Quantification}
\newacronym{V&V}{V\&V}{Verification and Validation}
\newacronym{LEVM}{LEVM}{Linear Eddy Viscosity Models}
\newacronym{PCS}{PCS}{Principal Coordinate System}
\newacronym{DNS}{DNS}{Direct Numerical Simulation}
\newacronym{QoI}{QoI}{Quantities of Interest}
\newacronym{DUU}{DUU}{Design Under Uncertainty}

\def\CorrectionsReviewOne{2}     

\ifnum\CorrectionsReviewOne=1    
    \newcommand{\cancelText}[1]{{\textcolor{red}{\sout{#1}}}}%
    \newcommand{\newText}[1]{{\textcolor{blue}{#1}}}%
\fi

\makeatletter
\ifnum\CorrectionsReviewOne=2    
    \newcommand{\cancelText}[1]{\@bsphack\@esphack}%
    \newcommand{\newText}[1]{#1}%
\fi

\def\CorrectionsReviewTwo{2}     

\ifnum\CorrectionsReviewTwo=1    
    \newcommand{\cancelTextTwo}[1]{{\textcolor{red}{\sout{#1}}}}%
    \newcommand{\newTextTwo}[1]{{\textcolor{blue}{#1}}}%
\fi

\makeatletter
\ifnum\CorrectionsReviewTwo=2    
    \newcommand{\cancelTextTwo}[1]{\@bsphack\@esphack}%
    \newcommand{\newTextTwo}[1]{#1}%
\fi

\def\plotPNGorPDF{1}     

\ifnum\plotPNGorPDF=1    
    
\fi

\makeatletter
\ifnum\plotPNGorPDF=2    
    
\fi

\makeatletter
\def\@email#1#2{%
 \endgroup
 \patchcmd{\titleblock@produce}
  {\frontmatter@RRAPformat}
  {\frontmatter@RRAPformat{\produce@RRAP{*#1\href{mailto:#2}{#2}}}\frontmatter@RRAPformat}
  {}{}
}%

\makeatother
\begin{document}

\preprint{AIP/Physics of Fluids}

\title[Physically constrained uncertainty estimation]{Physically constrained eigenspace perturbation for turbulence model uncertainty estimation}
\author{Marcel Matha*}
\author{Christian Morsbach}%
 \email{marcel.matha@dlr.de.}
\affiliation{ 
\gls{DLR}, Linder Höhe, 51147 Cologne, Germany}


\date{\today}

\begin{abstract}
Aerospace design is increasingly incorporating Design Under Uncertainty based approaches to lead to more robust and reliable optimal designs. These approaches require dependable estimates of uncertainty in simulations for their success. The key contributor of predictive uncertainty in Computational Fluid Dynamics (CFD) simulations of turbulent flows are the structural limitations of Reynolds-averaged Navier-Stokes models, termed model-form uncertainty. Currently, the common procedure to estimate turbulence model-form uncertainty is the Eigenspace Perturbation Framework (EPF), involving perturbations to the modeled Reynolds \cancelTextTwo{S}\newTextTwo{s}tress tensor within physical limits. The EPF has been applied with success in design and analysis tasks in numerous prior works from the industry and academia. Owing to its rapid success and adoption in several commercial and open-source CFD solvers, in depth Verification and Validation of the EPF is critical. In this work, we show that under certain conditions, the perturbations in the EPF can lead to Reynolds stress dynamics that are not physically realizable. This analysis enables us to propose a set of necessary physics-based constraints, leading to a realizable EPF. We apply this constrained procedure to the illustrative test case of a converging-diverging channel, and we demonstrate that these constraints limit physically implausible dynamics of the Reynolds stress tensor, while enhancing the accuracy and stability of the uncertainty estimation procedure. 

\end{abstract}

\maketitle

\section{\label{sec:introduction} Introduction \protect}
As computational resources continue to advance, the aerospace industry is experiencing a notable increase in the degree of digitization, leading to faster design cycles with the help of \gls{CFD}. 
In order to accelerate the optimization of designs and streamline virtual certification procedures, numerical approximations of the \gls{RANS} equations is a judicious choice. This choice not only upholds an acceptable level of fidelity but also computational efficiency for its purposes in design.
However, the \gls{RANS} equations necessitate the modeling of the second-moment Reynolds stress tensor $\boldmath{\tau}$.
Closure models, commonly referred to as turbulence models, attempt to express \gls{QoI} \cancelTextTwo{Interest} that are not measured like the Reynolds \cancelTextTwo{S}\newTextTwo{s}tresses as a function of measured quantities like the local mean rate of strain.
While turbulence modeling offers practicality and facilitates efficient simulations, it also imposes inherent limitations in achieving high levels of accuracy.
Moreover, the assumptions made in the functional representation of turbulence models introduce model-form (epistemic) uncertainties as soon as their applicability range is exceeded \newText{\cite{Duraisamy, XIAO20191}}. 
This is particularly relevant for complex engineering flows such as the ones encountered in turbomachinery components.
To provide a few examples, the prediction accuracy of common \gls{LEVM} turbulence models suffers in flows characterized by adverse pressure gradient, separation and reattachment, surface curvature\cancelText{, etc.} \newText{and secondary flow.} 
Due to the definition of the Reynolds stresses in \gls{LEVM} (see also introduction of Boussinesq approximation in \cref{sec:Method}), the tensor only carries information on the mean rate of strain, hence the model is unable to account for rotational effects and streamline curvature~\cite{Speziale, CRAFT1996108}. Additionally, in the isotropic eddy viscosity hypothesis, excluding representation of any anisotropic normal Reynolds stresses hinders the accurate consideration of secondary flow~\cite{Mompean}.\\
Accounting for the inherent uncertainties in simulations is key towards robust designs. 
That is why approaches to quantify the uncertainties associated with turbulence closure models play an important role, especially in industrial applications with turbulent flows.
The only approach capable of addressing the epistemic uncertainty inherent in turbulence closure modeling is the \gls{EPF} that was initially proposed by Emory~\cite{Emory2011}.  
This methodology builds upon the limited functional relationship of the Reynolds stresses.
Selective perturbation of the Reynolds stress tensor within physically bounds combined with sampling from the resulting \gls{CFD} solutions is an innovative model-form \gls{UQ} approach~\cite{Emory2013, Gorle2013, EmoryThesis, iaccarino2017eigenspace}.
The underlying modeling structure of the tensor perturbation involves perturbations in both eigenvalues and eigenvectors, which is comprehensively described in \cref{sec:Method}.
These perturbations can be interpreted as altering the shape and the orientation of the Reynolds stress tensor ellipsoid \cite{iaccarino2017eigenspace, mishra2019theoretical, Matha2023}.
Due to its unique characteristics and persuasive interpretability of its simulation outcomes, the \gls{EPF} has been used in various engineering applications \cite{garcia2014quantifying, EmoryTurbo, mishra2017uncertainty, Gorle2019, Lamberti, cook2019optimization, razaaly2019optimization, mishra2020design, Mukhopadhaya, Hornshoj, EidiDataFree, Thompson2019, Chu2022, gori2022confidence, Gori2022Springer}.
For this reason, the ability of perturbing the eigenspace of the Reynolds stresses has been integrated into numerous \gls{CFD} solvers~\cite{Edeling, MishraSU2, Gorle2019, MathaCF}.
In addition to that, the emergence of machine learning strategies guided the path towards data-driven enhancements of the \gls{EPF}~\cite{Edeling, heyse2021estimating, Eidi2022, MathaCF}. \\
As there is the need for \gls{V&V} of novel \gls{CFD} methods, this paper addresses the underlying modeling rationale of this framework. 
Recently, we have already proposed a novel advancement in the context of the \gls{EPF}, that focuses on ensuring realizable Reynolds stresses and consistency between the envisioned conceptual and the implemented computational model~\cite{Matha2023}.
While the theoretical modeling structure and limitations of the eigenvalue perturbation have been exhaustively discussed~\cite{mishra2019theoretical}, the analysis of the eigenvector perturbation remains incomplete so far. \\
In this article, we undertake a detailed examination of the foundation and ramifications of the eigenvector perturbations.
This thorough analysis of the Reynolds stress tensor's eigenvector perturbation in the context of \gls{RANS} equations, enables us to show that eigenvector perturbation, as they are currently implemented, may lead to non-realizable Reynolds stress tensor dynamics. 
Moreover, we highlight numerical stability issues that may arise as a consequence, potentially preventing broader application of this approach. 
Therefore, we derive and propose a novel idea to prevent implausible Reynolds stress tensor dynamics in the current paper.

\section{\label{sec:Method} Accounting for turbulence modeling uncertainty using the tensor perturbation framework\protect}

Despite the ongoing increase in computational resources, solving the set of Navier-Stokes equations for turbulent flows by scale-resolving simulations in the design phase for industrially relevant devices operating at high Reynolds numbers cannot be expected in the near future.
As engineers and system designers are rather interested in rapid iteration cycles, the ability to make decisions based on statistical consideration of the mean flow is still industrial practice. 
Hence, all flow quantities can be split into a mean and a fluctuating part, according to $\phi = \overline{\phi}+\phi'$.
To accommodate this need for compressible flows, a density weighted average (Favre-average) is performed, whereby 
\begin{equation}
\phi = \widetilde{\phi}+\phi'' \ \text{and} \ \ \overline{\rho}
\widetilde{\phi} = \overline{\rho \phi}
\end{equation}
holds for all instantaneous quantities except density $\rho$ and pressure $p$.
In the scope of this paper, we will use the term \gls{RANS} for the favre-averaged Navier-Stokes equations, although \textit{Reynolds-averaging} was initially developed for incompressible flows.
The statistically Favre-averaged momentum equation following Einstein's notation convention
\begin{equation}
\label{eq:RANSmomentum}
    \frac{\partial}{\partial t}\left(\overline{\rho} \widetilde{u_i}\right)+\frac{\partial}{\partial x_j}\left(\overline{\rho} \widetilde{u_j} \widetilde{u_i}\right) = -\frac{\partial \overline{p}}{\partial x_i} + \frac{\partial}{\partial x_j} \left(\sigma_{ij} - \overline{\rho} \widetilde{u_i' u_j'}\right) 
\end{equation}
describes the change of the mean momentum in both time and space, attributed to acting mean forces such as pressure gradients and divergence of viscous stresses (for the sake of simplicity, gravitational forces and forces due to rotating frames of reference are neglected).\\
Note: To shorten and simplify the notation, we denote the mean velocities by a capital letter $\widetilde{u_i}\rightarrow U_i$ and omit the overline for density $\overline{\rho}\rightarrow \rho$ and pressure $\overline{p}\rightarrow p$. Additionally we use ${x,y,z}$ for ${x_1, x_2, x_3}$ in the following.\\
Based on Stokes' hypothesis the mean viscous stresses $\boldmath{\sigma}$ depend on the strain-rate tensor $S_{ij}=\frac{1}{2}\left(\frac{\partial U_i}{\partial x_j}+\frac{\partial U_j}{\partial x_i}\right)$ and kinematic viscosity denoted as $\nu$:
\begin{equation}
\label{eq:laminarStresses}
\sigma_{ij} = 2 \rho \nu \left(S_{ij}-\frac{1}{3}S_{kk} \delta_{ij}\right)
\end{equation}
In addition to these stresses, the right hand side of the equation contains unknown correlations of fluctuating velocities $\tau_{ij} = \widetilde{u_i'' u_j''}$, called the turbulent stresses or Reynolds stresses~\cite{Pope}.
To close the set of equations and facilitate computational simulations, there exist numerous approximation methods.
A widely used modeling assumption is the representation of Reynolds stresses as an isotropic function of the scalar eddy viscosity $\nu_t$ and the mean rate of strain tensor, drawing an analogy to the representation of viscous stresses
\begin{equation}
\label{eq:boussinesq}
    \tau_{ij} = -2 \nu_t \left(S_{ij} - \frac{1}{3}S_{kk}\delta_{ij}\right) + \frac{2}{3} k \delta_{ij} \ \text{.}
\end{equation}
The equation mentioned above, also known as the Boussinesq approximation, ensures, that the trace of the resulting tensor is twice the turbulent kinetic energy $k = \frac{1}{2} \tau_{kk}$.
State-of-the-art two-equation turbulence models, such as Menter's SST $k-\omega$ model~\cite{Menter}, typically solve additional partial differential transport equations for the turbulent kinetic energy and the turbulent dissipation rate and reconstruct the eddy viscosity afterwards to close the set of equations.
The assumed linear relationship between Reynolds stresses and strain-rate tensor, however, is not universally valid, as already discussed in~\cref{sec:introduction}.
Consequently, any simulation using the Boussinesq assumption contains inherent epistemic uncertainty. 
The perturbation of Reynolds stress tensor's eigenspace~\cite{Emory2011} is the method of choice in order to account for turbulence modeling uncertainty on \gls{QoI}. 
The underlying methodology is described in the following\cancelTextTwo{ section}.

The symmetric, positive semi-definite Reynolds stress tensor $\tau_{ij}$ can be decomposed into an anisotropy tensor $a_{ij}$ and an isotropic component 
\begin{equation}
	\label{eq:anisotropy}
		\tau_{ij} = k \left(a_{ij} + \frac{2}{3}\delta_{ij}\right) \ \text{.}
\end{equation}
Eddy viscosity based turbulence models assume that the tensorial characteristics of the anisotropy tensor are solely dictated by the mean rate of strain tensor (see \cref{eq:boussinesq})
\begin{equation}
	\label{eq:anisotropy2}
		a_{ij} = - 2 \frac{\nu_t}{k} \left(S_{ij} - \frac{1}{3}S_{kk}\delta_{ij}\right) \ \text{.}
\end{equation}
The epistemic discrepancy in the evaluation of Reynolds stresses can be represented by the tensor $Q_{ij}$, such that the true Reynolds stresses are
\begin{equation}
\label{eq:uncertaintytensor}
\begin{split}
    \tau^{\text{true}}_{ij} &=  \tau^{\text{modeled}}_{ij}+ Q_{ij} \\
    &= -2 \nu_t \left(S_{ij} - \frac{1}{3}\frac{\partial u_k}{\partial x_k}\delta_{ij}\right) + \frac{2}{3} k \delta_{ij} + Q_{ij} \\
    &= k a_{ij} + \frac{2}{3} k \delta_{ij} + Q_{ij}\ \text{.}
\end{split}
\end{equation}
Building upon the concept of the eigenspace perturbation approach, the structural uncertainty of the Reynolds stress tensor can be split into contributors of shape, alignment and amplitude of the tensor. 
Therefore, the anisotropy tensor can be represented by an eigenspace decomposition
\begin{equation}
	\label{eq:spectralDecompositionAnisotropy}
		a_{ij} = v_{in} \Lambda_{nl} v_{jl} \ \text{.}
\end{equation}
The orthonormal eigenvectors form the matrix $v_{in}$ while the traceless diagonal matrix $\Lambda_{nl}$ contains the corresponding ordered eigenvalues $\lambda_k$.
When Boussinesq approximation is used, the eigenvectors of the anisotropy tensor coincide with those of the strain-rate tensor, while the eigenvalues $\lambda_k$ are solely dependent on the strain-rate tensor's eigenvalues $\gamma_k$ and its trace 
\begin{equation}
\label{eq:eigenvaluesStrain}
\lambda_{k} = -2\frac{\nu_t}{k} \left(\gamma_k -\frac{S_{kk}}{3}\right)\ \text{.}
\end{equation}
Evidently, the Reynolds stress tensor features identical eigenvectors as well, however the eigenvalues of the Reynolds stress tensor are 
\begin{equation}
\label{eq:eigenvaluesReStress}
\psi_k = k (\lambda_k + 2/3) \ \text{.}
\end{equation} 
Inserting \cref{eq:spectralDecompositionAnisotropy} into \cref{eq:uncertaintytensor} leads to:
\begin{equation}
\label{eq:uncertaintytensor2}
\begin{split}
    \tau^{\text{true}}_{ij} &= k \left(v_{in} \Lambda_{nl} v_{jl}\right) + \frac{2}{3} k \delta_{ij} + Q_{ij}\ \text{.}
\end{split}
\end{equation}
Because of the tensorial properties, the tensor $Q_{ij}$ can be decomposed into 
\begin{equation}
\label{eq:uncertaintytensorDecomposition}
\begin{split}
    Q_{ij} = \Delta k \left(\Delta v_{in}^{a} \Delta  \Lambda_{nl}^{a} \Delta v_{jl}^{a}\right) + \frac{2}{3} \Delta k \delta_{ij} \text{,}
\end{split}
\end{equation}
whereby $\Delta$ describe the error terms for turbulent kinetic energy (amplitude), alignment (eigenvectors) and shape (eigenvalues).

As precisely quantifying the uncertainty of the turbulence model in representing the modeled Reynolds stress tensor is a challenging task, the developers and founders of the methodology rather try to estimate the uncertainty by sampling from possible solution space.
Hence, it is not the aim to apply a correct Reynolds stress tensor $\tau^{\text{true}}_{ij}$ but a perturbed, physical\newTextTwo{ly}\cancelTextTwo{-}realizable one, which is called $\tau^*_{ij}$ \cite{MathaCF}.
Following the line of argument above, the \gls{EPF}, considered in this work, creates a perturbed state of the Reynolds stress tensor defined as
\begin{equation}
	\label{eq:spectralDecompositionR*}
	\begin{split}
	    \tau_{ij}^* &= k \left(a_{ij}^* + \frac{2}{3}\delta_{ij}\right) \\
		            &= k \left(v_{in}^* \Lambda_{nl}^* v_{jl}^* + \frac{2}{3}\delta_{ij}\right) \ \text{,}
	\end{split}
\end{equation}
where $a_{ij}^*$ indicates the perturbed anisotropy tensor, $\Lambda_{nl}^*$ represents its perturbed eigenvalue matrix and $v^*_{in}$ is the perturbed eigenvector matrix.
Adhering to the procedure established in the majority of previously published works, there is no explicit modification of the turbulent kinetic $k$ energy.
Instead the level of turbulence is manipulated indirectly by altering the production of turbulence due to affirmative perturbations of eigenvalues and eigenvectors, as will be clarified in subsequent sections.

\begin{figure}
\centering
\begin{tikzpicture}
\node[] at (0,3) {
                \includegraphics[width=0.07\textwidth]{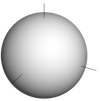}};
\node[] at (-2,-2.6) {
                \includegraphics[width=0.07\textwidth]{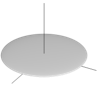}};
\node[] at (2,-2.6) {
                \includegraphics[width=0.07\textwidth]{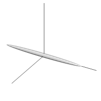}};
\draw [-](-2,-1.5) -- (2,-1.5) node[] {};
\draw [-](-2,-1.5) -- (0,1.964) node[] {};
\draw [-]( 2,-1.5) -- (0,1.964) node[] {};
\draw [gray, dashed]( -0.3, 0.5) -- (0,1.964) node[] {};
\draw [gray, dashed]( -1.34, -0.72) -- (-2,-1.5) node[] {};
\draw [gray, dashed]( -0.3, 0.5) -- (0.2,0.08) node[] {};
\draw [gray, dashed]( 0.82, -0.48) -- (2,-1.5) node[] {};
\draw [red]( -1.34, -0.72) -- (-0.3, 0.5) node[] {};
\filldraw[black] (-0.3, 0.5) circle (2pt) node[anchor=west]{$\mathbf{x}$};
\filldraw[black] (1.7, -1.5) circle (0.01pt) node[anchor=north west]{$\mathbf{x}_{\mathrm{1C}}$};
\filldraw[black] (-1.7, -1.5) circle (0.01pt) node[anchor=north east]{$\mathbf{x}_{\mathrm{2C}}$};
\filldraw[black] (0, 1.964) circle (0.01pt) node[anchor=south]{$\mathbf{x}_{\mathrm{3C}}$};
\filldraw[black] ( -1.34, -0.72) circle (2pt) node[anchor=west]{$\mathbf{x}^*$};
\node[red] at (0, -0.3) {$\Delta_B = \frac{\mathbf{x}^* - \mathbf{x}}{\mathbf{x}_{\mathrm{2C}} - \mathbf{x}}$};
\node[black] at (0, -1.7) {\footnotesize Two-component};
\node[black, rotate=60] at (-1.19, 0.3) {\footnotesize Axisymmetric};
\node[black, rotate=-60] at (1.19, 0.3) {\footnotesize Axisymmetric};
\end{tikzpicture}
\vspace{-0.5cm}
\caption{Systematic representation of the eigenvalue perturbation  within the barycentric triangle and its effect on the shape of the Reynolds stress tensor ellipsoid.}
\label{fig:baryCentricTriangle}
\end{figure}
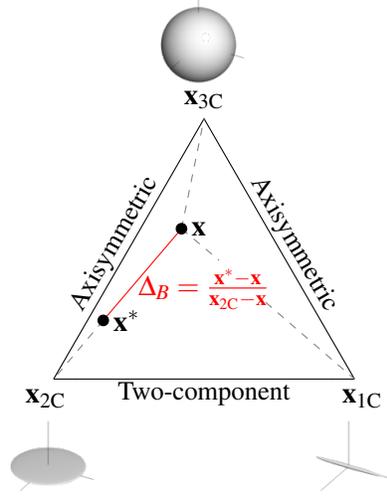

\newTextTwo{\subsection{\label{sec:eigenvaluePerturbation} Eigenvalue perturbation}}
As the components of the symmetric anisotropy tensor are bounded according to the realizability constraints~\cite{Schumann1977}, the respective eigenvalues can be transformed into barycentric coordinates~\cite{Banerjee2007}.
By defining the vertices $\mathbf{x}_{\mathrm{1C}}, \mathbf{x}_{\mathrm{2C}}, \mathbf{x}_{\mathrm{3C}}$ of an equilateral triangle, representing the componentiality of turbulence (three-component, isotropic limit (3C), two-component axisymmetric limit (2C) and the one-component limit (1C))~\cite{Terentiev2006}, the mapping from anisotropy eigenvalues to barycentric coordinates is defined as
\begin{equation}
\label{eq:barycentricMapping}
    \begin{split}
	   \mathbf{x} =& \frac{1}{2}\mathbf{x}_{\mathrm{1C}}\left(\lambda_1-\lambda_2\right)+ \mathbf{x}_{\mathrm{2C}}\left(\lambda_2-\lambda_3\right)+ \mathbf{x}_{\mathrm{3C}} \left(\frac{3}{2}  \lambda_3+1\right)  \\
        \mathbf{x} =& \mathbf{B} \boldmath{\lambda} 
        \quad \text{with} \ \lambda_1\geq\lambda_2 \geq\lambda_3 \ \text{,}
     \end{split}
\end{equation}
The envisioned perturbation of the eigenvalues of the anisotropy tensor within physically permissible limits is grounded on shifting the barycentric state within the borders of the barycentric triangle~\cite{Emory2011}. Using the pseudoinverse of $\mathbf{B}$, any perturbed eigenvalues are expressed through remapping 
\begin{equation}
	\label{eq:perturbedEigenvalues}
		\boldsymbol{\lambda}^* = \mathbf{B}^{+} \mathbf{x}^* \ \text{,}
\end{equation}
where the relocated position $\mathbf{x}^*$ results from linear interpolation between starting point $\mathbf{x}$ and target point $\mathbf{x}_{(t)} \in \left\{ \mathbf{x}_{\mathrm{1C}}, \mathbf{x}_{\mathrm{2C}}, \mathbf{x}_{\mathrm{3C}} \right\}$
\begin{equation}
	\label{eq:perturbationMagnitude}
		\mathbf{x}^* = \mathbf{x} + \Delta_B \left(\mathbf{x}_{(t)} -\mathbf{x}\right) \ \text{.}
\end{equation}
The relative distance $\Delta_B \in [0, 1]$ controls the magnitude of eigenvalue perturbation as presented in~\cref{fig:baryCentricTriangle}.
Traditional eddy viscosity-based turbulence models assume that this eddy viscosity is a scalar, known as the isotropic eddy viscosity. Thus, turbulence behaves as an isotropic medium. The eigenvalue perturbation modulates this to an orthotropic medium, where turbulence behaves differently along each eigen-direction~\cite{mishra2019theoretical}, accounting for the sensitivity of the model with respect to the anisotropic characteristics of turbulence.

\newTextTwo{\subsection{\label{sec:eigenvectorPerturbation} Eigenvector perturbation}}

 Given that \gls{LEVM} rely on the Boussinesq approximation in~\cref{eq:boussinesq}, the Reynolds stress, the anisotropy and the strain-rate tensor share identical eigendirections, as already discussed in~\cref{sec:Method}. However, this relationship results in inaccuracies in predicting certain flows, e.g. involving flow separation and reattachment.
 Nevertheless, even for simple turbulent boundary layer flow there is significant misalignment between scale-resolving simulation (such as \gls{DNS}) and \gls{RANS} model predicted eigenvectors of the Reynolds stress tensor~\cite{Matha2021Neurips}.
 Hence, the eigenspace perturbation idea adds a perturbation to the eigenvectors. In contrast to the eigenvalues, there are no actual bounds for the orientation of the Reynolds stress tensor ellipsoid. To address this issue, Iaccarino et al.~\cite{iaccarino2017eigenspace} suggest to make use of the boundedness of the Frobenius inner product of the Reynolds stress and the strain-rate tensor, called the turbulent production $P_k$ of the turbulent kinetic energy transport equation.
Based on the relationship of the strain-rate and Reynolds stress tensor for \gls{LEVM} (see~\cref{eq:boussinesq}), the bounds of the turbulent production term can be written in terms of their eigenvalues $\psi_i$ and $\gamma_i$ \cite{Lasserre}:
\begin{equation}
\label{eq:turbulentProduction}
        P_k =-\tau_{ij}\frac{\partial U_i}{\partial x_j} 
            \in \left[\psi_1\gamma_3+\psi_2\gamma_2+\psi_3\gamma_1, \ \psi_1\gamma_1+\psi_2\gamma_2+\psi_3\gamma_3\right]  \ \text{.}
\end{equation}
As the Reynolds stress and the strain-rate tensor share the same eigenvectors for \gls{LEVM}, the lower bound of the turbulent production term can be obtained by commuting the first and third eigenvector of the Reynolds stress tensor that manipulates the relationship between eigenvalues and respective eigendirections.
The permutation of first and third eigenvector results in a reconstructed Reynolds stress tensor based on \cref{eq:spectralDecompositionR*}, which is equivalent to the one obtained by rotating the eigenvector matrix $\mathbf{v}$ around the second eigenvector by $\pi/2$, see \cref{app:permutingEigenvectors}.
Whereas keeping the ordering of eigenvectors in case of \gls{LEVM}, evidently leads to the upper limit of the turbulent production.
In the subsequent section, we outline, why the eigenvector perturbation can lead to implausible dynamics of the Reynolds stress tensor combined with an unrealistically derived turbulent production term.

\section{\label{sec:realizablity}Adherence to realizable Reynolds stress tensors and realizable Reynolds stress tensor dynamics}
\subsection{Insights from the Reynolds stress tensor's eigenspace perturbation and implications for turbulent boundary layers}
The significant advantage of the \gls{EPF} lies in its ability to generate a perturbed and realizable Reynolds stress tensor from an unperturbed one.
This is accomplished by ensuring that the realizability condition, saying that the Reynolds stress tensor must be positive semi-definite, is met~\cite{Schumann1977}.
To illustrate, when perturbing the eigenvalues of the modeled Reynolds stress tensor, choosing $\Delta \leq 1$ (see \cref{fig:baryCentricTriangle}) inevitably leads to fulfillment of the realizability condition as the perturbed Reynolds stress anisotropy eigenvalues remain inside the barycentric triangle.
Recently, we addressed an appropriate way to incorporate eigenvector perturbations in a self-consistent manner in order to obtain the desired realizable Reynolds stresses~\cite{Matha2023}.
However, while the current formulation of the realizability principle is valuable, it is not comprehensive or adequate in ensuring that the evolution of the Reynolds stress, from one physically permissible state to another, remains physically plausible. 
Indeed, under certain conditions, the realizable Reynolds stress tensor, obtained through eigenspace perturbation, may become physically implausible leading to turbulent stress dynamics, which are rather not realizable.

\begin{figure}[bt]
 \centering
 \begin{tikzpicture}
 \node[] at (0,0){
                \includegraphics[width=0.1\linewidth]{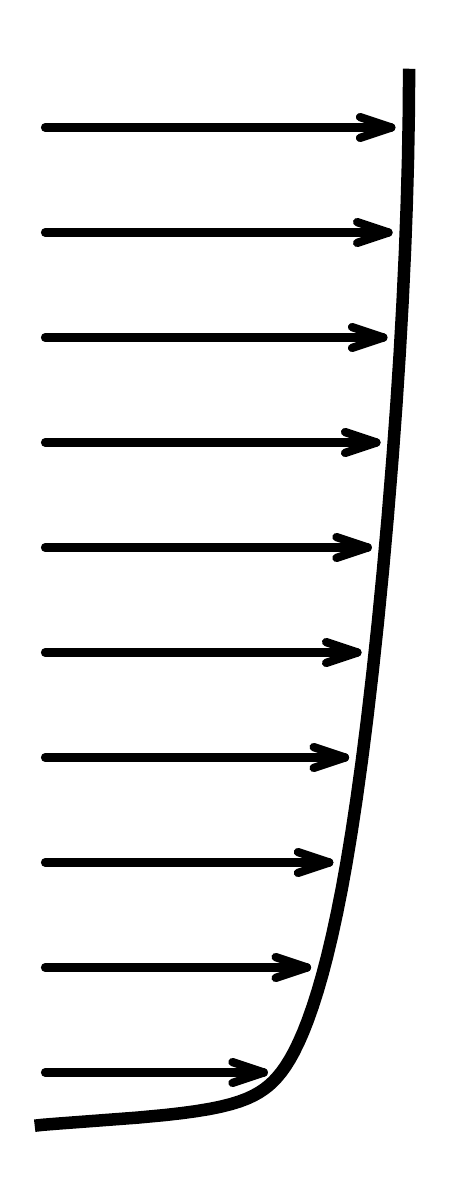}};
\draw [-{Stealth[length=3mm, width=3mm]}] (-0.9,-2.2) -- (1.1,-2.2);
\draw [-{Stealth[length=3mm, width=3mm]}] (-0.9,-2.2) -- (-0.9,2.3);
\node[black] at (0.7, -2.5) {\footnotesize$x$};
\node[black] at (-1.2, 1.82) {\footnotesize$y$};
\node[black] at (1.4, 0.6) {\footnotesize $U_1 \geq 0$};
\node[black] at (1.4, -0.2) {\footnotesize $\frac{\partial U_1}{\partial y} \geq 0$};
\node[black] at (1.4, -1.0) {\footnotesize $\sigma_{12} \geq 0$};
\node[black] at (1.4, -1.8) {\footnotesize $\tau_{12} \leq 0$};
\end{tikzpicture}
\caption{\label{fig:1DboundaryLayerSketch}Schematics of steady, fully developed 1D boundary layer flow}
\end{figure}
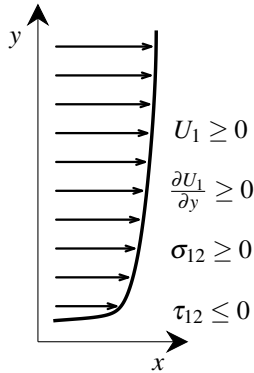

An exemplary case to illustrate these conditions is the turbulent boundary layer, whereby we consider the flow to be steady, 1D and fully developed. This is equivalent to analyzing half of a symmetric infinite channel flow, as sketched in \cref{fig:1DboundaryLayerSketch}.
Hence, by setting $\frac{\partial}{\partial t}=0$, $U_2=U_3=0$, $\frac{\partial \boxed{}}{\partial x} =  \frac{\partial \boxed{}}{\partial z}=0$ (except $\frac{\partial p}{\partial x} \neq 0$), \cref{eq:RANSmomentum} simplifies to
\begin{equation}
\label{eq:1Dbl_momentum}
    \frac{\partial p}{\partial x} =  \frac{\partial}{\partial y} \left(\sigma_{12} - \rho \tau_{12}\right) \ \text{.}
\end{equation}
The diffusion based on viscous stresses has to be balanced by a source term, associated with a streamwise pressure gradient.
Applying the isotropic eddy viscosity assumption (see \cref{eq:boussinesq}), the Reynolds stress tensor for 1D boundary layer flow becomes
\begin{equation}
\label{eq:ReStress1D}
\boldmath{\tau}= 
\begin{pmatrix}
\frac{2}{3}k & -\nu_t \frac{\partial U_1}{\partial y} & 0\\
-\nu_t \frac{\partial U_1}{\partial y} & \frac{2}{3}k & 0\\
0 & 0 & \frac{2}{3}k 
\end{pmatrix} \ \text{.}
\end{equation}
The eigenvalues $\psi_1 = \frac{2}{3}k +\nu_t \frac{\partial U_1}{\partial y}$, $\psi_2 = \frac{2}{3}k$, $\psi_3 = \frac{2}{3}k -\nu_t \frac{\partial U_1}{\partial y}$ come along with the respective eigenvectors $\mathbf{v_1} = \left(\frac{-1}{\sqrt{2}}, \frac{1}{\sqrt{2}}, 0\right)^T$, $\mathbf{v_2} = \left(0, 0, 1\right)^T$ and $\mathbf{v_3} = \left(\frac{1}{\sqrt{2}}, \frac{1}{\sqrt{2}}, 0\right)^T$.
By means of the eigenspace decomposition $\tau_{ij} = v_{in} \Psi_{nl} v_{jl}$ and employing the eigenvector matrix $\mathbf{v}= (\mathbf{v_1}, \mathbf{v_2}, \mathbf{v_3})$, the shear stress component of the Reynolds stress tensor can be reformulated
\begin{equation}
\label{eq:shearStress1D}
\begin{split}
\tau_{12} &= v_{11}v_{12}\psi_1+v_{21}v_{22}\psi_2+v_{31}v_{32}\psi_3\\
\end{split}
\end{equation}
Inserting the unperturbed eigenvectors and eigenvalues of the Reynolds stress tensor, as outlined above, into \cref{eq:shearStress1D} results in a strictly negative Reynolds shear stress component $\tau_{12} = \frac{1}{2} \left(\psi_3-\psi_1\right)$ as $\psi_1 \geq \psi_3$.
However, when perturbing the eigenspace orientation according to the approach of Iaccarino et al.~\cite{iaccarino2017eigenspace}, we obtain $\tau^*_{12} = \frac{1}{2} \left(\psi_1-\psi_3\right) \geq 0$ as $\psi_1 \geq \psi_3$.
In conclusion, the simple permutation of first and third eigenvector leads to different sign of the relevant shear stress shaping the boundary layer profile.
At first glance, this already seems to violate obvious flow physics.

Nevertheless, we aim to present a conceptual explanation for this phenomenon. 
In order to qualitatively assess the physical relationships related to a change in sign of the Reynolds shear stress, we insert Stokes' hypothesis (see \cref{eq:laminarStresses}) and the eddy viscosity hypothesis (see \cref{eq:boussinesq}) into \cref{eq:1Dbl_momentum}
\begin{equation}
\label{eq:1Dbl_momentum_2}
\begin{split}
    \frac{\partial p}{\partial x}  &=  \frac{\partial}{\partial y} \left(\rho \left(\nu+  \nu_t\right) \frac{\partial U_1}{\partial y} \right) \ \text{.}
\end{split}
\end{equation}
Consequently, a change in sign of Reynolds stress component $\tau_{12}$ would equate to an effective negative turbulent eddy viscosity $\nu_t$. 
Negative eddy viscosity means that momentum flux from regions of higher momentum is transported to regions of lower momentum. 
This implies a countergradient transport that is physically implausible. Additionally, this phenomenon is associated with positively correlated Reynolds stresses and mean velocity gradients.
Following the example of steady, fully developed 1D boundary layer, the turbulent production term $P_k =-\tau_{ij}\frac{\partial U_i}{\partial x_j} = - \tau_{12} \frac{\partial U_1}{\partial y}$ will be negative, if $\tau_{12}$ becomes positive because of eigenvector permutation.
Such negative turbulent production denotes transferring energy from the turbulent scales to the mean kinetic energy, a process that is deemed physically implausible as well~\cite{Gayen2011}.

\begin{figure}[bt]
\captionsetup[subfloat]{captionskip=-4pt}
\centering
    \begin{subfloat}[\label{fig:ProductionNoEigenvaluePerturbation}]
         {\includegraphics[width=0.3\textwidth]{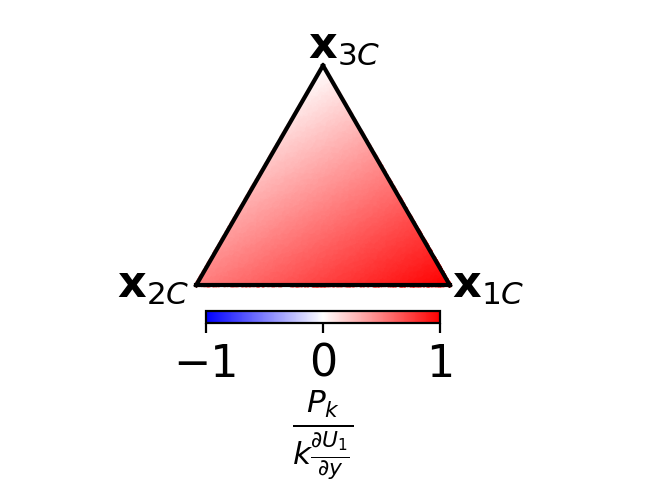}}
    \end{subfloat}
    \begin{subfloat}[\label{fig:ProductionEigenvaluePerturbation}]
         {\includegraphics[width=0.3\textwidth]{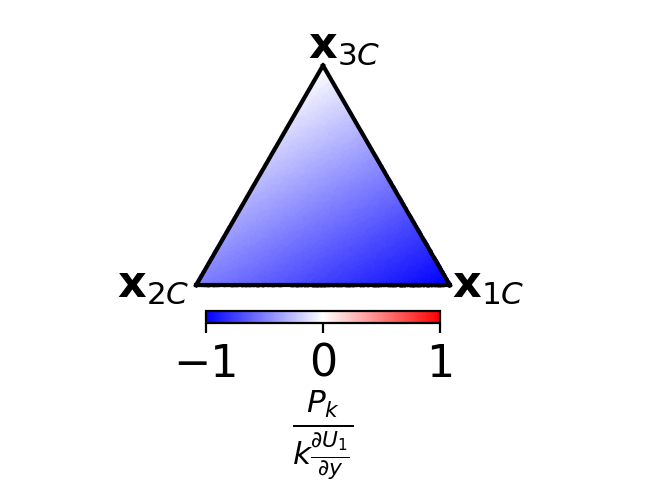}}
    \end{subfloat}
\vspace{-0.2cm}
\caption{Comparison of the effect of eigenspace perturbation on the turbulent production term $P_k$ in case of fully developed boundary layer flow. Effect of pure eigenvalue perturbation is shown in (a), while (b) presents the effect, when combining the permutation of the eigenvectors $\mathbf{v_1}$ and $\mathbf{v_3}$ and eigenvalue perturbation within the barycentric triangle.}
\label{fig:effectOfEigenvaluePerturbationOnProduction}
\end{figure}

In contrast to the eigenvector perturbation, a pure eigenvalue perturbation is incapable of inducing a change in the sign of $\tau_{12}$ for fully developed boundary layer flow.
Indeed, \cref{fig:ProductionEigenvaluePerturbation} additionally serves to illustrate the observation, that applying eigenvector perturbation lead to negative turbulent production and, consequently, negative effective eddy viscosity for any eigenvalue perturbation that falls within the bounds of the barycentric triangle.
As depicted in \cref{fig:effectOfEigenvaluePerturbationOnProduction}, the absolute value of the turbulent production reaches its maximum at the one-component limiting state of turbulence, whereas it becomes zero for an isotropic Reynolds stress tensor, which is in accordance to the finding of Gorlé et al.~\cite{Gorle2019}.
This illustrative example demonstrates that, in the context of wall-bounded, boundary layer like flows, the suggested eigenvector permutation of first and third eigenvector can give rise to non-realizable Reynolds stress tensor dynamics in the set of \gls{RANS} equations. 
Therefore, there is the need for a physics-based constraint that ensures not only realizable Reynolds stresses but also plausible Reynolds stress tensor dynamics. Subsequently, we derive this constraint, verify its validity and suggest its future usage within the \gls{EPF}.

\subsection{\label{sec:derivatrionOfConstrain}Simplified derivation of realizable eigenvector perturbation dynamics for wall-bounded flows}

\begin{figure}[tb]
\captionsetup[subfloat]{captionskip=-10pt}
\centering
    \begin{subfloat}[\label{fig:tensorRotationAlpha}]
        {\begin{tikzpicture}
            \node[] at (-3,0) {
                \includegraphics[width=0.3\textwidth, trim=2.1cm 7.5cm 1cm 7cm, clip=True]{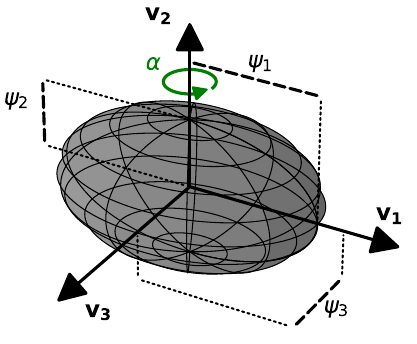}};
            \node[] at (2,0) {
                \includegraphics[width=0.3\textwidth, trim=2.5cm 7.5cm 1cm 7cm, clip=True]{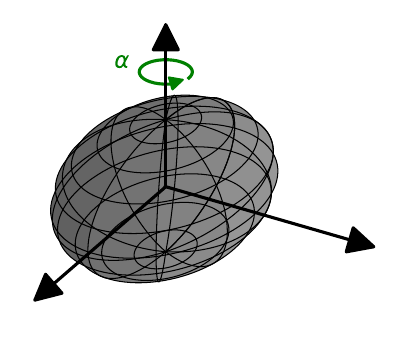}};
            \draw[->, ultra thick] (-1.3,0.7) 
                arc [
                    start angle=140,
                    end angle=40,
                    x radius=1cm,
                    y radius =1cm
                    ] ;
        \end{tikzpicture}}  
    \end{subfloat}
    \begin{subfloat}[\label{fig:1DEulerAngle}]
        {\includegraphics[width=0.3\textwidth]{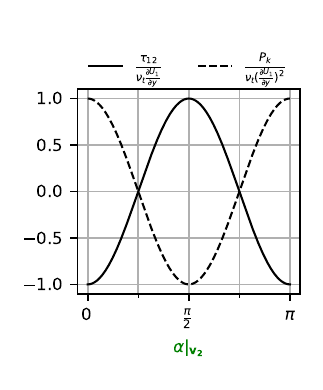}}
    \end{subfloat}
\vspace{-0.2cm}
\caption{Rotation of the eigenvector matrix of the Reynolds stress tensor around second eigenvector $\mathbf{v_2}$ by $\alpha$. The schemtical impact of the rotation on the Reynolds stress tensor ellipsoid is shown in (a). (b) shows the effect of eigenvector rotation on the Reynolds shear stress component and the turbulent production. This plot is created based on assuming 1D boundary layer flow, as sketched in \cref{fig:1DboundaryLayerSketch}. The eigenvectors of $\tau_{ij}$ presented in \cref{eq:ReStress1D} are rotated by $\alpha$. The resulting $\tau_{12}$ and $P_k$ (see \cref{eq:turbulentProduction}) are evaluated subsequently.}
\label{fig:exploringEffectofEigenvectorPermutation}
\end{figure}

As the second eigenvector of the Reynolds stress tensor in \cref{eq:ReStress1D}, is $\mathbf{v_2} =\left(0,0,1\right)^T$, the rotation matrix for any rotation around this eigenvector simplifies to
\begin{equation}
\mathbf{R_z}= 
\begin{pmatrix}
\cos\left(\alpha\right) &  -\sin\left(\alpha\right) & 0\\
\sin\left(\alpha\right) & \cos\left(\alpha\right) & 0\\
0 &0 & 1 
\end{pmatrix}\ \text{,}
\end{equation}
(choosing $\alpha = \pi/2$ results in Iaccarino's permutation of first and third eigenvector~\cite{iaccarino2017eigenspace} see \cref{app:permutingEigenvectors}).
The general rotation of the Reynolds stress tensor ellipsoid around its second eigenvector is sketched in ~\cref{fig:tensorRotationAlpha}.

The objective is to derive a condition that evidently causes a change of sign for the shear Reynolds stress component $\tau_{12}$, ultimately resulting in non-realizable Reynolds stress tensor dynamics.
Therefore, we formulate the rotated eigenvector matrix based on the unperturbed eigenvector matrix $\mathbf{v}$\cancelTextTwo{listed in \cref{sec:realizablity}}
\begin{equation}
\begin{split}
\mathbf{v^*}&= \mathbf{R_z} \mathbf{v}\\
&= \frac{1}{\sqrt{2}}\begin{pmatrix}
-\sin\left(\alpha\right) - \cos\left(\alpha\right) & 0 &  \cos\left(\alpha\right)-\sin\left(\alpha\right)\\
-\sin\left(\alpha\right) + \cos\left(\alpha\right) & 0 & \cos\left(\alpha\right) + \sin\left(\alpha\right)\\
0 & \sqrt{2} & 0 
\end{pmatrix}
\end{split}
\end{equation}
Hence, the resulting Reynolds shear stress based on \cref{eq:shearStress1D} becomes:
\begin{equation}
\label{eq:shearStress1DRotatedConstrain}
\begin{split}
\tau_{12}^* 
= \frac{1}{2}\left(\psi_3-\psi_1\right) \cos\left(2\alpha\right)  \mathop{=}\limits^! 0
\end{split}
\end{equation}
Consequently, \cref{eq:shearStress1DRotatedConstrain} holds true for isotropic turbulence, as $\psi_1=\psi_3$ and any rotation angle $\alpha = \frac{\pi}{2} n - \frac{\pi}{4}$ with $n \in \mathbb{N}$.
The relationship of the rotation angle and the shear stress component is verified by a step-by-step analysis presented in \cref{fig:1DEulerAngle}. The resulting dependency of the Reynolds shear stress component is exactly the analytically derived one in \cref{eq:shearStress1DRotatedConstrain}.\\
Note: The rotation of the Reynolds stress tensor is symmetric to $\pi/2$, which means that any rotation around $\pi/2-\delta$ results in the same tensor as any rotation around $\pi/2+\delta$.
Therefore, $\alpha = \pi/4$ is the appropriate choice as the smallest angle at which a sign change occurs.

The mean of the cosine in \cref{fig:1DEulerAngle} has to be zero in order to obtain zero crossing of the turbulent production at exactly $\alpha =\pi/4$. 
In other words, it is required that the maximum and the minimum value of the turbulent production have equal absolute magnitude but opposite signs. 
Equating the lower and upper bound of the inner Frobenius product \cref{eq:turbulentProduction} leads to
\begin{equation}
\label{eq:conditionForZeroProduction}
\begin{split}
    -P_{k_{min}} &= P_{k_{max}} \\
    -\psi_1\gamma_3-\psi_2\gamma_2-\psi_3\gamma_1 &= \psi_1\gamma_1+\psi_2\gamma_2+\psi_3\gamma_3\\
    \psi_1 \left(\gamma_1+\gamma_3\right)+2\psi_2\gamma_2+\psi_3 \left(\gamma_1+\gamma_3\right)&=0 \ \text{.}
\end{split}
\end{equation}
Thus, rotating the orthogonal eigenvectors around the second eigenvector by an angle of $\pi/4$, results in zero turbulent production if $\gamma_1=-\gamma_3$ and $\gamma_2 = 0$. 
This conditions always holds true for fully-developed 1D boundary layers, as there is only a single velocity gradient present in the flow. 
However, any 2D flow featuring vanishing divergence of the velocity field does also satisfy \cref{eq:conditionForZeroProduction}. 
This means, that $\frac{\partial U_1}{\partial x} = -\frac{\partial U_2}{\partial y}$, given that $U_3$ is the vanishing velocity component.

\subsection{A posteriori validation of the suggested constraint on the eigenvector perturbation}
\begin{figure}[tb]
 \centering
\includegraphics[width=0.55\linewidth]{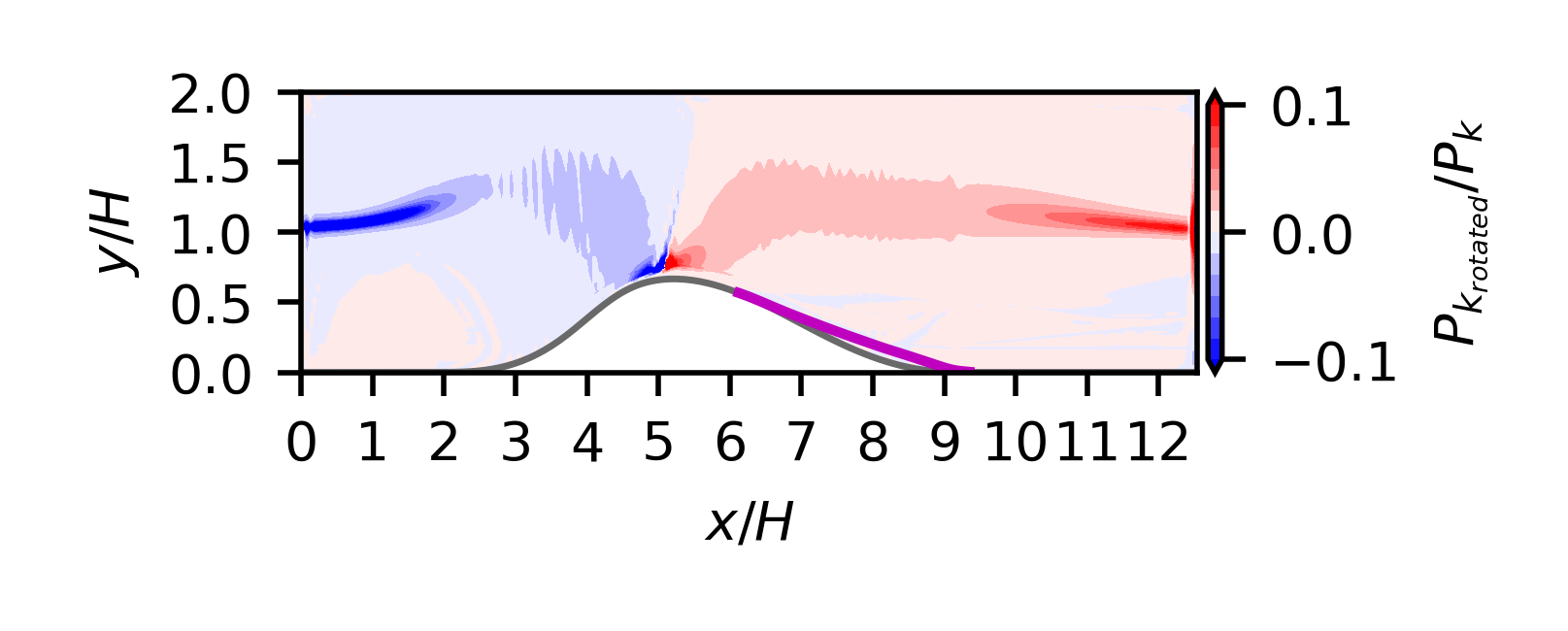}
\vspace{-1cm}
\caption{\label{fig:productionRotatedReStress} Distribution of turbulent production term $P_{k_{\text{rotated}}}$, when rotating the eigenvectors of the Reynolds stress tensor around second eigenvector by $\alpha =\pi/4$. For better interpretability the resulting production is scaled by the unperturbed turbulent production $P_k$. The magenta line indicates $U_1 =0$}
\end{figure}

To corroborate our findings, we analyse another generic flow scenario, which is the 2D converging-diverging channel flow \cite{Laval}\cancelTextTwo{(see schematics of the test case in \cref{app:convDivSketch})}.
\newTextTwo{This test case consists of viscous walls with and without curvature as sketched in \cref{fig:convDivSketch}a.}
\newTextTwo{Our numerical setup uses inlet boundary conditions extracted from a fully developed turbulent boundary layer at $Re_{\tau} = 617$.}
\newTextTwo{The derived mass flow rate is forced using a boundary controller, which adjusts the static pressure at the outlet of the computational domain.
We conducted a \gls{RANS} grid independence study using a low-Reynolds resolution ($y^+ \leq 1$) at solid walls and by applying Menter SST $k-\omega$ \gls{LEVM}\cite{Menter}.}
Based on\cancelTextTwo{a previously performed \gls{RANS} simulation using the Menter SST $k-\omega$ \gls{LEVM}\cite{Menter}} \newTextTwo{this}, we conduct analysis in post processing \newTextTwo{for the finest mesh featuring a resolution of 242x242x1 grid points (see \cref{fig:convDivSketch}b)}.
The resulting velocity gradients, the eddy viscosity and the turbulent kinetic energy are used to determine the Reynolds stress tensor following Boussinesq's approximation (see \cref{eq:boussinesq}).
According to our derivation above, the eigenvectors of these Reynolds stress tensors are rotated around the second eigenvector by $\alpha = \pi/4$ in the entire domain as a first step. The rotated Reynolds stress tensors are composed using \cref{eq:spectralDecompositionR*}.
Subsequently, we can compare the resulting turbulent production term (see ~\cref{eq:turbulentProduction}) after rotating the eigenvectors with the one based on the initial Reynolds stress tensor.
The comparison, presented in \cref{fig:productionRotatedReStress}, reveals a reduction in the effective turbulent production due to the rotation as expected. This observation confirms the exemplarily derived relationship of the turbulent production term with respect to eigenvector rotation of the Reynolds stress tensor.
\begin{figure}[tb]
\centering
 \includegraphics[width=0.55\linewidth]
{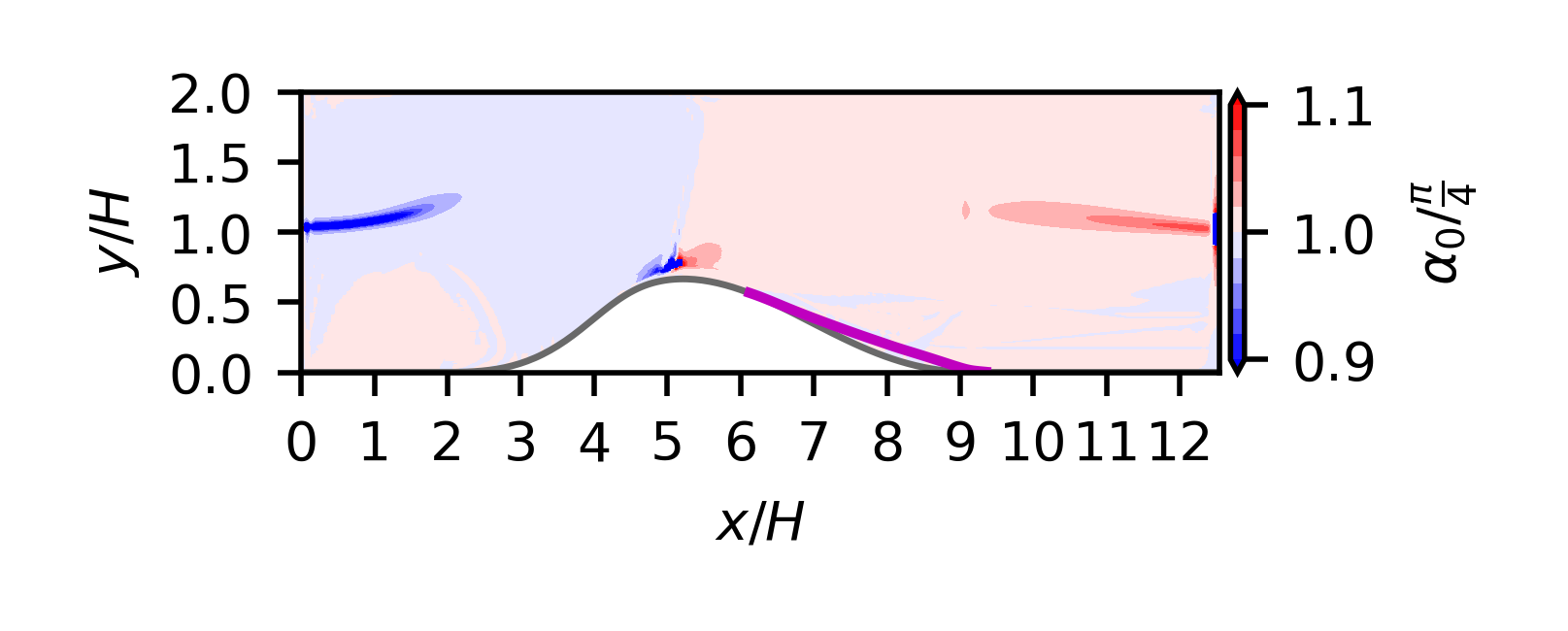}
\vspace{-1cm}
\caption{Rotation around second eigenvector of the Reynolds stress tensor by $\alpha_0$ leads to zero turbulent production. In order to better classify the discrepancy from $\pi/4$, the determined angle $\alpha_0$ is presented as a fraction of $\pi/4$. The magenta line indicates $U_1 =0$}
\label{fig:zeroProductionAngle}
\end{figure}
As a second step, we further validate the derivations by solving an optimization problem for achieving zero turbulent production by an eigenvector rotation of the Reynolds stress tensor given the velocity gradients of the previously performed \gls{RANS} simulation of the 2D converging-diverging channel. \Cref{fig:zeroProductionAngle} shows the appropriate rotation angle $\alpha_0$ relative to $\pi/4$ that would lead to zero turbulent production term. 
The deviations from to the derived $\alpha$ under idealised conditions of a 1D boundary layer flow, can be ascribed to the fact, that the flow is not fully divergence-free in the outer parts of the boundary layers. 
However, as the optimized $\alpha_0$ differs by only 10\% at maximum from $\pi/4$, we believe, that restricting the eigenvector rotation of the Reynolds stress tensor to $\pi/4$ is a reasonable choice also for more complex flows.

To sum up, we have shown through mathematical analysis, that a simple eigenvector perturbation involving permuting the first and third eigenvector, may lead to implausible, not realizable Reynolds stress tensor dynamics. Based on that, a constraint that facilitates physically meaningful Reynolds stress tensor perturbations with respect to rotation of the eigenspace has been derived for wall-bounded, boundary layer like flows. Additionally, we have substantiated the derivations by presenting illustrative proofs.
In the next section, we apply the proposed eigenvector rotations in the \gls{EPF} \cancelTextTwo{implemented in our in-house \gls{CFD} solver \textit{TRACE}}.

\section{\label{sec:application}Application of physically constrained eigenvector perturbation}

\newTextTwo{The idea of the \gls{EPF} is to sample from possible solution space for certain \gls{QoI} attributed to perturbing the Reynolds stress tensor within the discussed physical bounds.
The entire framework was implemented in \gls{DLR}'s in-house \gls{CFD} solver \textit{TRACE}~\cite{Geiser19}.
\textit{TRACE} is a parallel Navier-Stokes flow solver that has been developed at \gls{DLR}'s Institute of
Propulsion Technology in close cooperation with MTU Aero Engines AG.
In the present work, we use the finite-volume method to discretize the compressible \gls{RANS} equations.
The \gls{EPF} and can be subdivided in several steps within each pseudo-time step of steady simulations:
\begin{enumerate}
    \item determine anisotropy tensor (see Equation \cref{eq:anisotropy})
    \item perturb eigenvalues of anisotropy tensor by choosing the relative perturbation strength $\Delta_B$ (cf. \cref{eq:perturbedEigenvalues} and \cref{eq:perturbationMagnitude})
    \item perturb eigenvectors of anisotropy (Reynolds stress) tensor by choosing the rotation angle $\alpha$. The rotation is done by rotating the eigenvector matrix around the second eigenvector $\mathbf{v^*} = \mathbf{R} \mathbf{v}$ with $\mathbf{R}$ according to \cref{eq:arbitraryRotationMatrix}.
    \item reconstruct perturbed Reynolds stress tensor $\tau_{ij}^*$ according to \cref{eq:spectralDecompositionR*}
    \item update of the viscous fluxes using $\tau_{ij}^*$ 
    \item update of the turbulent production term $P_k=-\tau_{ij}^*\frac{\partial u_i}{\partial x_j}$
\end{enumerate}}
The simulations of the flow within a converging-diverging channel serve to exemplify the application of the \gls{EPF} \newTextTwo{and further validate the proposed constraint on the eigenvector perturbation}. \newTextTwo{We compare against \gls{DNS} data by Laval et al.~\cite{Laval}.}
\cancelTextTwo{The test case consists of viscous walls with and without curvature as sketched in \cref{app:convDivSketch}. Available \gls{DNS} data~\cite{Laval} highlight, that the flow separates at the lower curved wall due to the adverse pressure gradient. We perform \gls{RANS} simulations with consistent inlet boundary conditions extracted from a fully developed turbulent boundary layer at $Re_{\tau} = 617$. The derived mass flow rate is forced using a boundary controller, which adjusts the static pressure at the outlet of the computational domain.}
The two-equation, Menter SST $k-\omega$~\cite{Menter} \gls{LEVM} is chosen to be the baseline model in the present investigation. 
Hence, the uncertainty estimates presented subsequently based on the \gls{EPF} can be attributed to the structural uncertainties within this particular turbulence model. 
As the amount of considered structural uncertainty increases with increasing eigenvalue perturbation, the most conservative estimation of the modelling uncertainty is obtained by choosing $\Delta_B=1.0$.
Nevertheless, according to latest publications~\cite{MathaCF}, intense Reynolds stress tensor perturbations may cause numerical convergence issues.

Following the approach proposed in our previous work~\cite{Matha2023}, the relative perturbation magnitude with respect to the relative shift in barycentric coordinates $\Delta_B$ has to be adjusted as a consequence of the convergence issues.
In the present study, we seek to apply a $\Delta_B$ as large as possible by steps of 0.1.
Consequently, while the full Reynolds stress tensor perturbation could be used for the 2C and 1C corners, the perturbation towards the isotropic corner had to be adjusted by $\Delta_B < 1$, as approaching the isotropic state results in a reduction of turbulent kinetic energy production.

Although, we have just derived that the maximum eigenvector rotation angle has to be $\alpha \leq \pi/4$, this constraint is necessary but not sufficient for practical applications.
The eigenvector modification by applying $\alpha \leq \pi/4$ may result in states of the Reynolds stress tensor that are indeed realizable and physically plausible but still lead to numerical stability issues. 
Therefore, we iteratively decrease the rotation angle by fractions of 10\% with respect to the maximum value of $\pi/4$. 
Besides examining the overall residuals and convergence of the static outlet pressure (controlled to maintain the prescribed mass flow outlet boundary condition) of each simulation, we evaluate the evolution of the the streamwise velocity. Therefore, we record iterative data at $x/H \in [0.5, 1, 2, 3, 4, 5, 6, 7, 8, 9, 10, 11, 12]$ every 1000 iterations and evaluate the relative error (standard deviation divided by the mean) over the last 100 snapshots. In order to distinguish between an unacceptable unstable and an acceptable converged solution, we use a maximum tolerable relative error of 1.5\% in each considered location.
The numerically achievable perturbations leading to converged \gls{RANS} results for this study of the convergence-divergence channel flow are summarized in \cref{tab:conv-div_simulations}.
In order to verify, that the eigenvector perturbation proposed by Iaccarino et al.\cite{iaccarino2017eigenspace} leads to unstable \gls{CFD} simulations as a result of non-realizable Reynolds stress tensor dynamics, we have conducted one exemplary simulation, presented in \cref{app:failingPerturbation}, applying eigenvector permutation without any eigenvalue perturbation.

\begin{table}
  \caption{\label{tab:conv-div_simulations}Selected turbulent target state (componentiality), $\Delta_B$ for eigenvalue and $\alpha$ for eigenvector perturbation of Reynolds stress tensor perturbation of flow within converging-diverging channel.}
  \begin{ruledtabular}
  \begin{tabular}{c c  c c c c c}
    simulation & \#1&\#2&\#3&\#4&\#5 &\#6\\
    target turbulent state & 1C & 1C& 2C & 2C& 3C & 3C\\
     $\Delta_B$ & 1.0 & 1.0 & 1.0& 1.0 & 0.2 & 0.2\\
     $\alpha$ & 0.0 &$\pi/10$ &0.0& $\pi/8$ & 0.0 & $\pi/20$\\
    \end{tabular}
  \end{ruledtabular}
\end{table}

\begin{figure}[tb]
 \centering
\includegraphics[width=\textwidth]{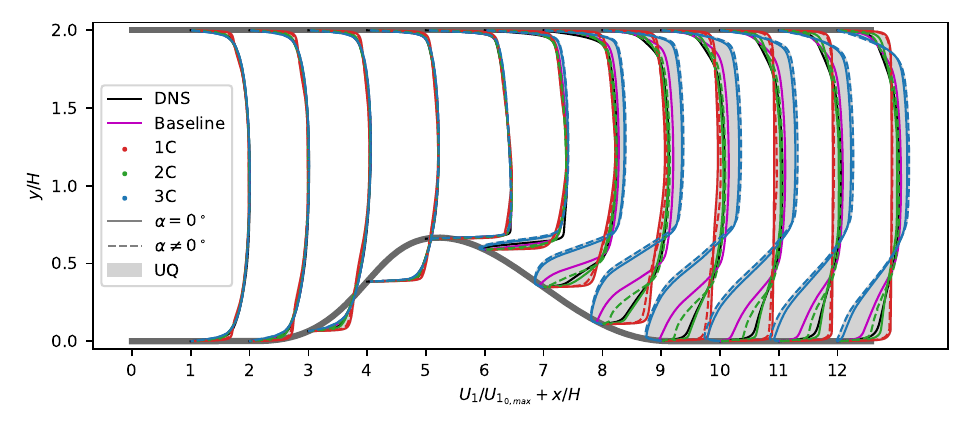}%
\vspace{-1cm}
\caption{\label{fig:uq_velocityX_conv_div} Estimated turbulence model uncertainty for the streamwise velocity inside of the converging-diverging channel based on the \gls{EPF}. $U_{1_{0\text{,max}}}$ is the maximum streamwise velocity of the baseline simulation at $x/H$=0. The settings for every eigenspace perturbation of the Reynolds stress tensor can be found in \cref{tab:conv-div_simulations}.}
\end{figure}
\begin{figure}[tb]
\captionsetup[subfloat]{captionskip=-10pt}
\centering
\begin{subfloat}[]
    {\includegraphics[width=0.495\textwidth]{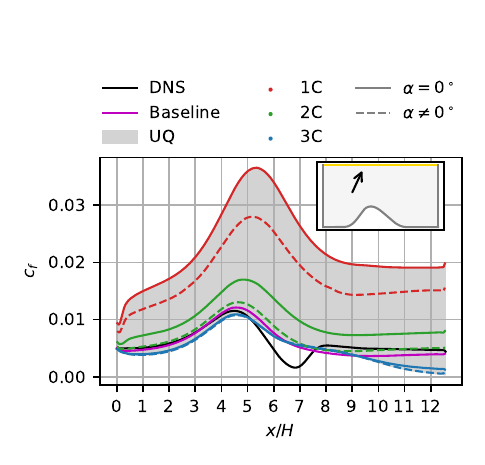}}
\end{subfloat}
\begin{subfloat}[]
    {\includegraphics[width=0.495\textwidth]{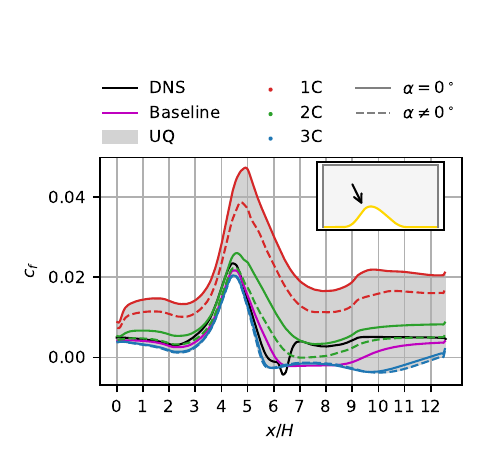}}
\end{subfloat}
\vspace{-1cm}
\caption{Turbulence model uncertainty based on the \gls{EPF} for the friction coefficient $c_f = \tau_w/\left(\frac{1}{2}\rho_0 U_{1_{0{\text{,max}}}}^2\right)$ at upper wall (a) and bottom wall (b) of the converging-diverging channel. The quantities with subscript $0$ indicate, that they are extracted at $x/H=0$. The settings for every eigenspace perturbation of the Reynolds stress tensor can be found in \cref{tab:conv-div_simulations}.}
\label{fig:surface_cf_conv_div}
\end{figure}
\begin{figure}[tb]
\captionsetup[subfloat]{captionskip=-10pt}
\centering
\begin{subfloat}[]
    {\includegraphics[width=0.495\textwidth]{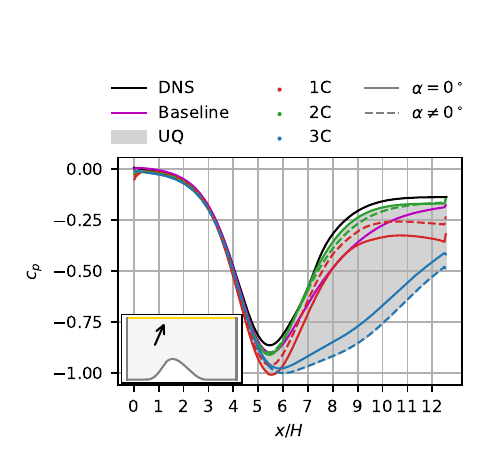}}
\end{subfloat}
\begin{subfloat}[]
    {\includegraphics[width=0.495\textwidth]{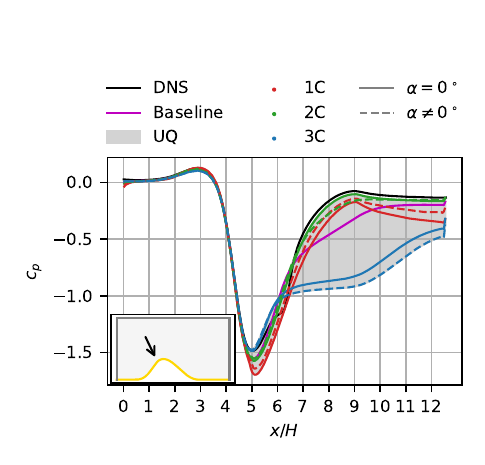}}
\end{subfloat}
\vspace{-1cm}
\caption{Turbulence model uncertainty based on the \gls{EPF} for the pressure coefficient $c_p = \left(p-p_0\right)/\left(\frac{1}{2}\rho_0 U_{1_{0{\text{,max}}}}^2\right)$ at upper wall (a) and bottom wall (b) of the converging-diverging channel. The quantities with subscript $0$ indicate, that they are extracted at $x/H=0$. The settings for every eigenspace perturbation of the Reynolds stress tensor can be found in \cref{tab:conv-div_simulations}.}
\label{fig:surface_cp_conv_div}
\end{figure}
\begin{figure}[!h]
\captionsetup[subfloat]{farskip=-5pt,captionskip=-15pt}
\centering
\begin{subfloat}[\gls{DNS}\cite{Laval}]
    {\includegraphics[width=0.495\textwidth]{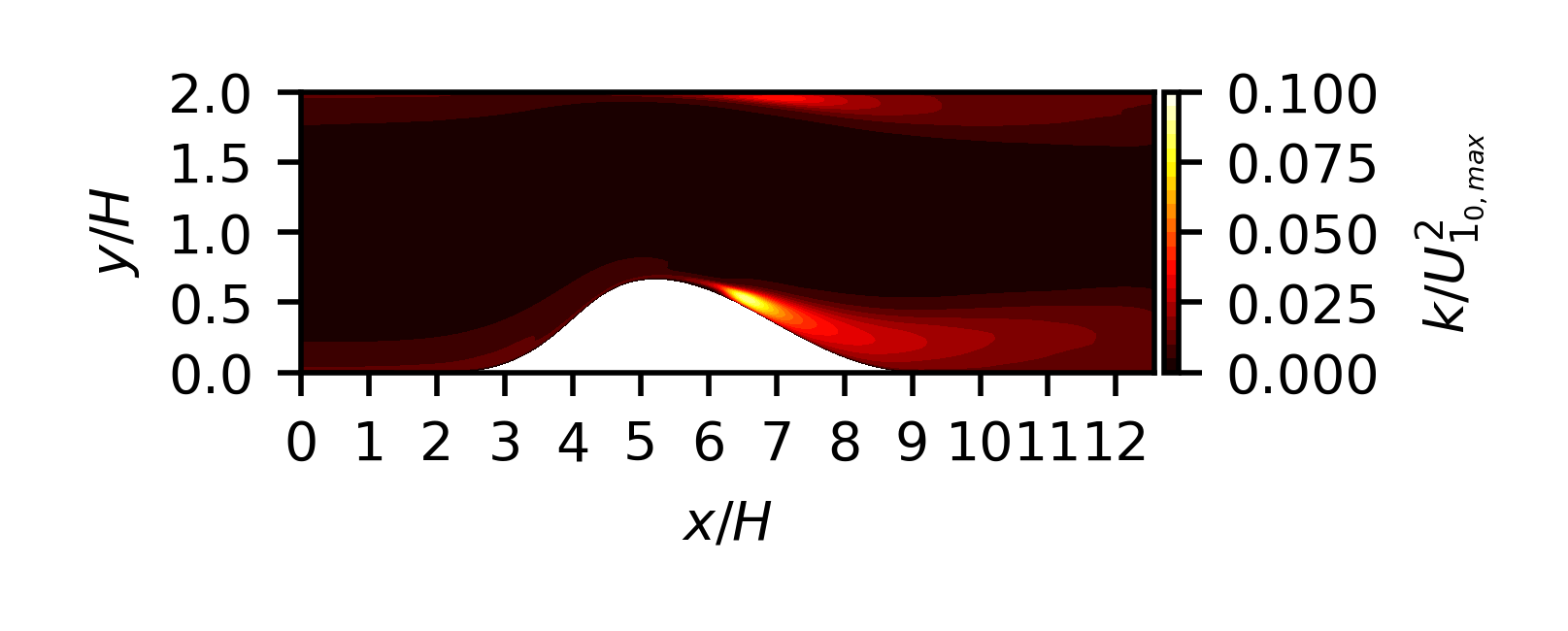}}
\end{subfloat}
\begin{subfloat}[\gls{RANS} Baseline]
    {\includegraphics[width=0.495\textwidth]{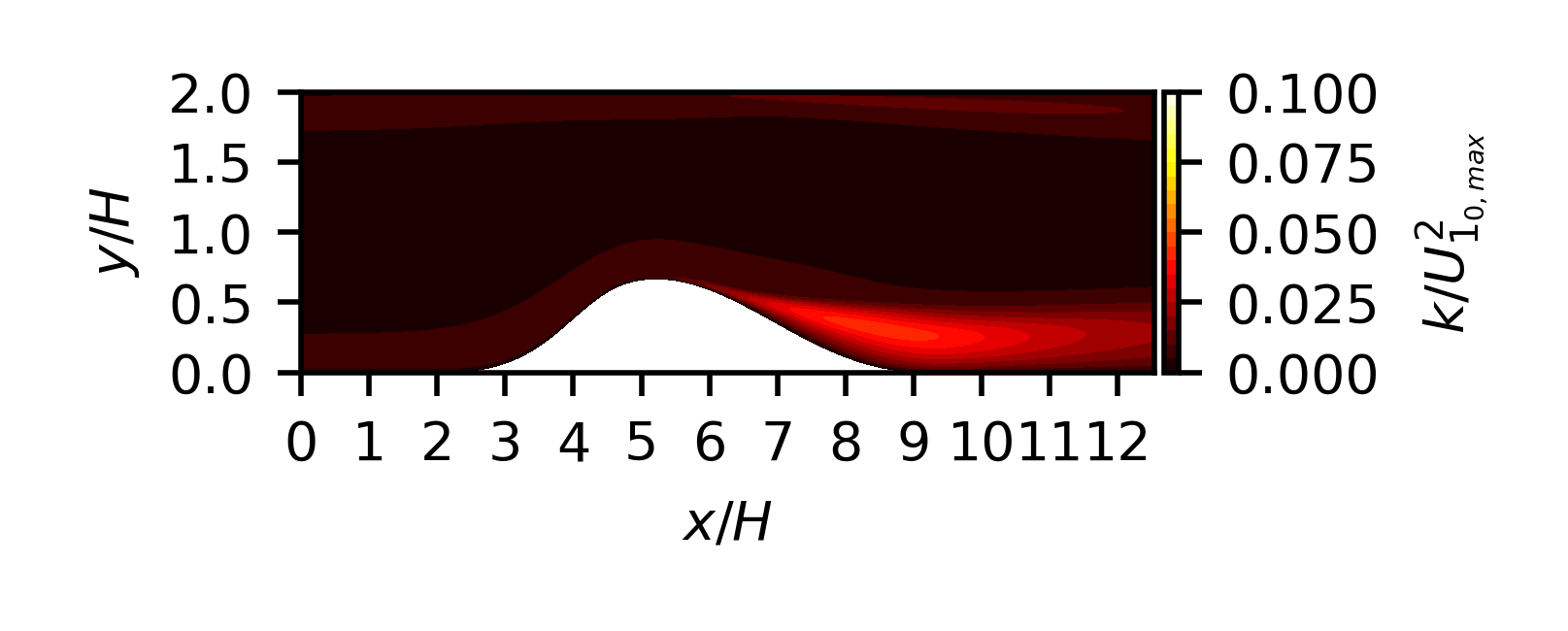}}
\end{subfloat}
\begin{subfloat}[Aiming for 1C without eigenvector rotation (\#1)]
    {\includegraphics[width=0.495\textwidth]{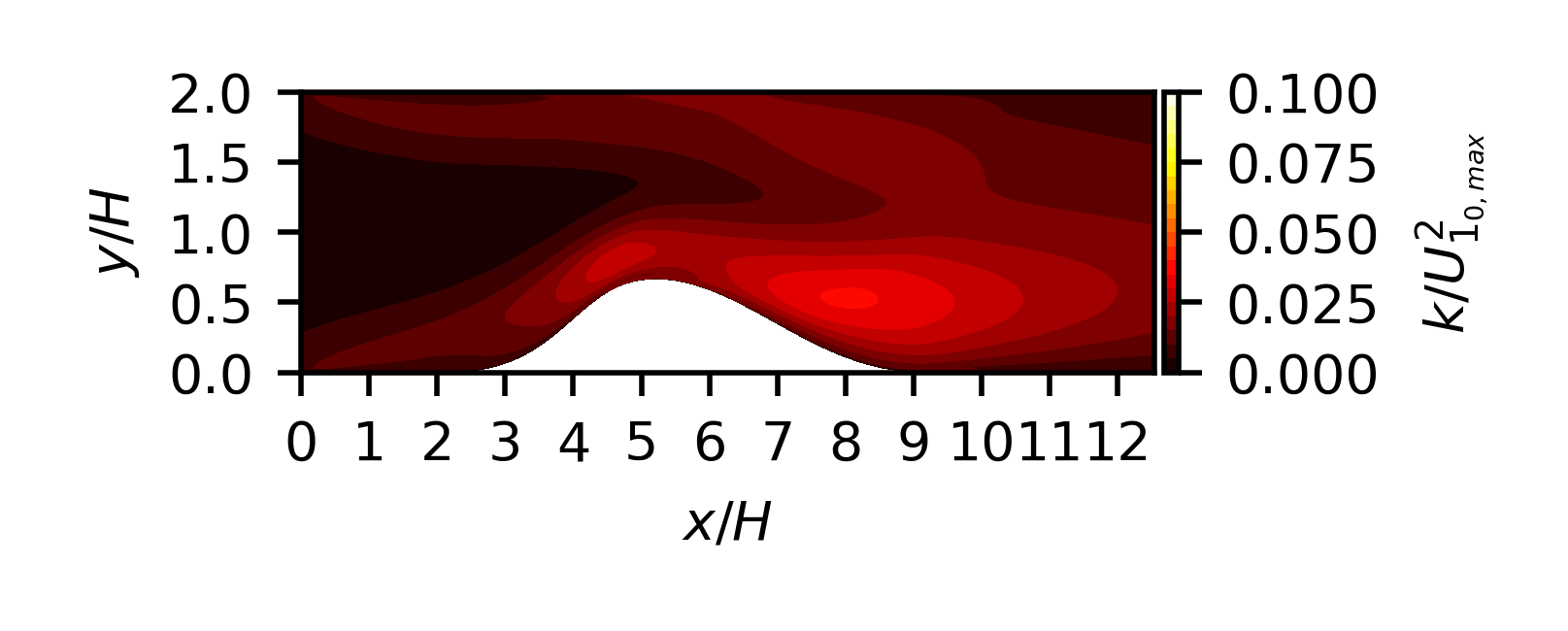}}
\end{subfloat}
\begin{subfloat}[Aiming for 1C with eigenvector rotation (\#2)]
    {\includegraphics[width=0.495\textwidth]{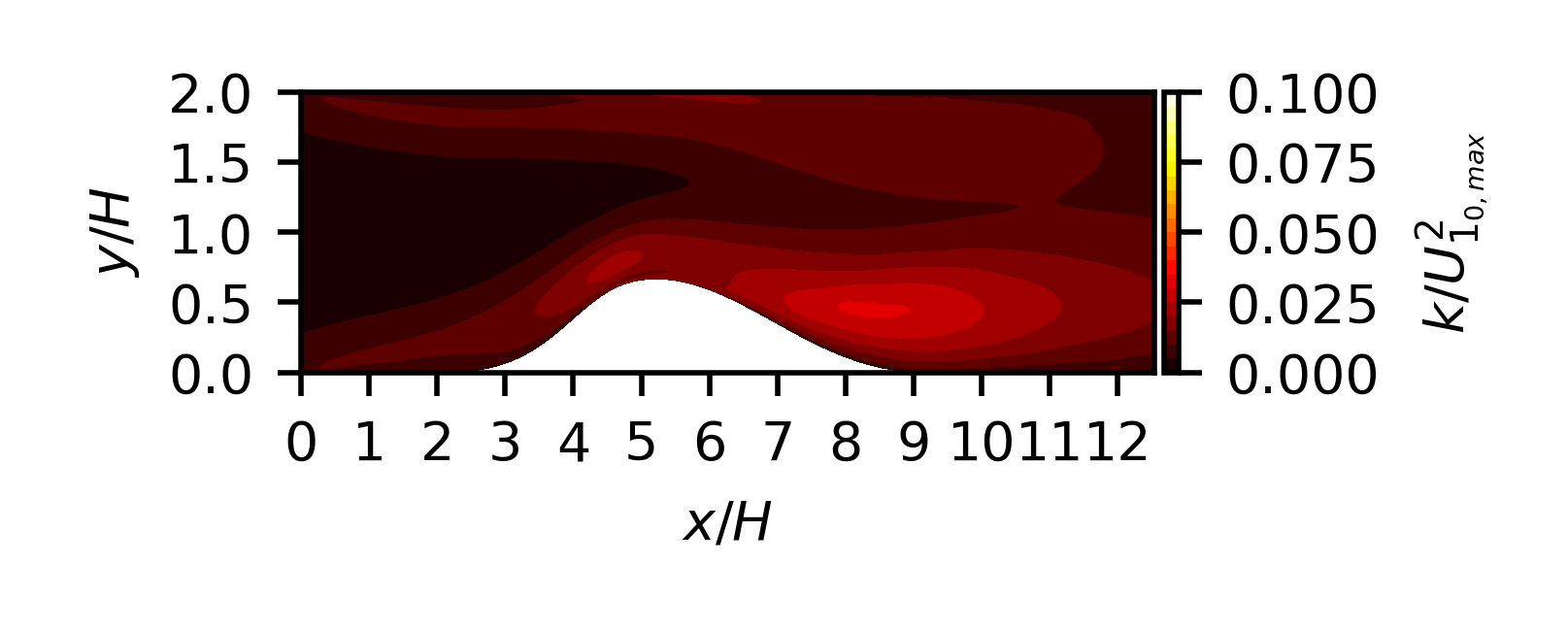}}
\end{subfloat}
\begin{subfloat}[Aiming for 2C without eigenvector rotation; (\#3)]
    {\includegraphics[width=0.495\textwidth]{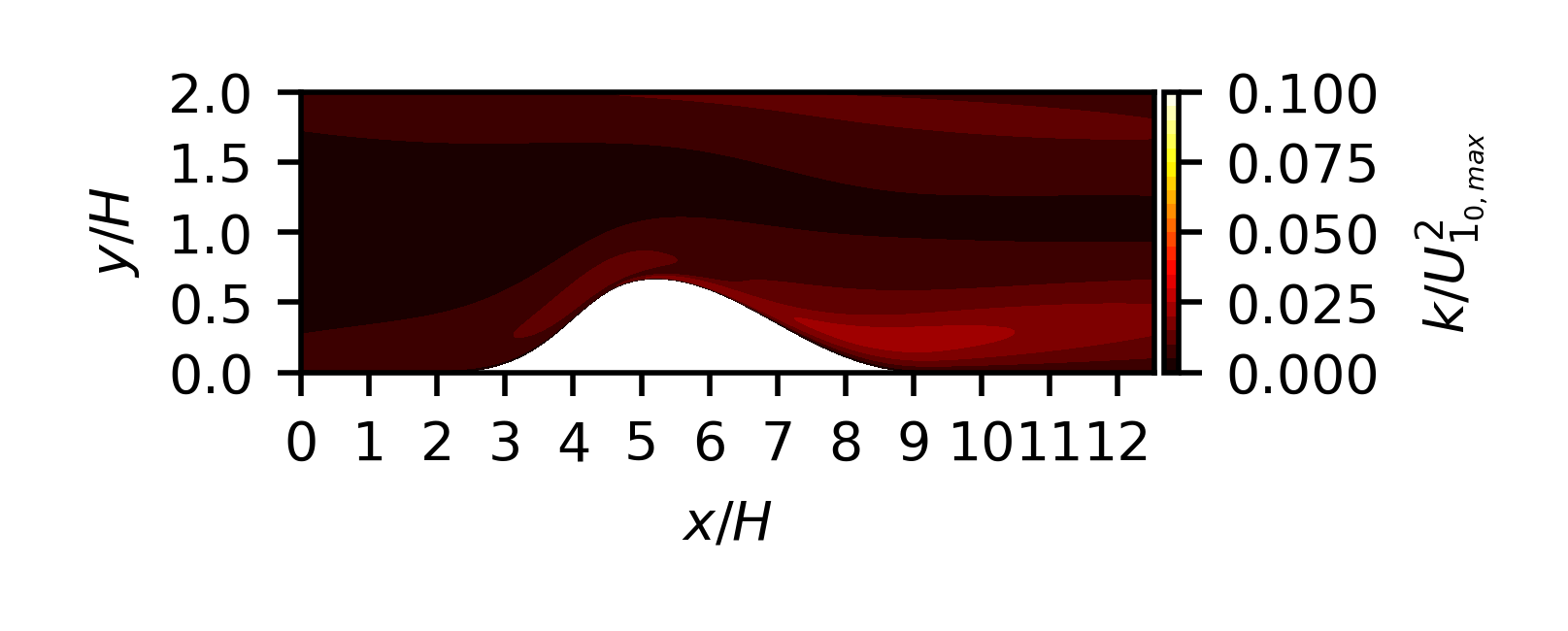}}
\end{subfloat}
\begin{subfloat}[Aiming for 2C with eigenvector rotation (\#4)]
    {\includegraphics[width=0.495\textwidth]{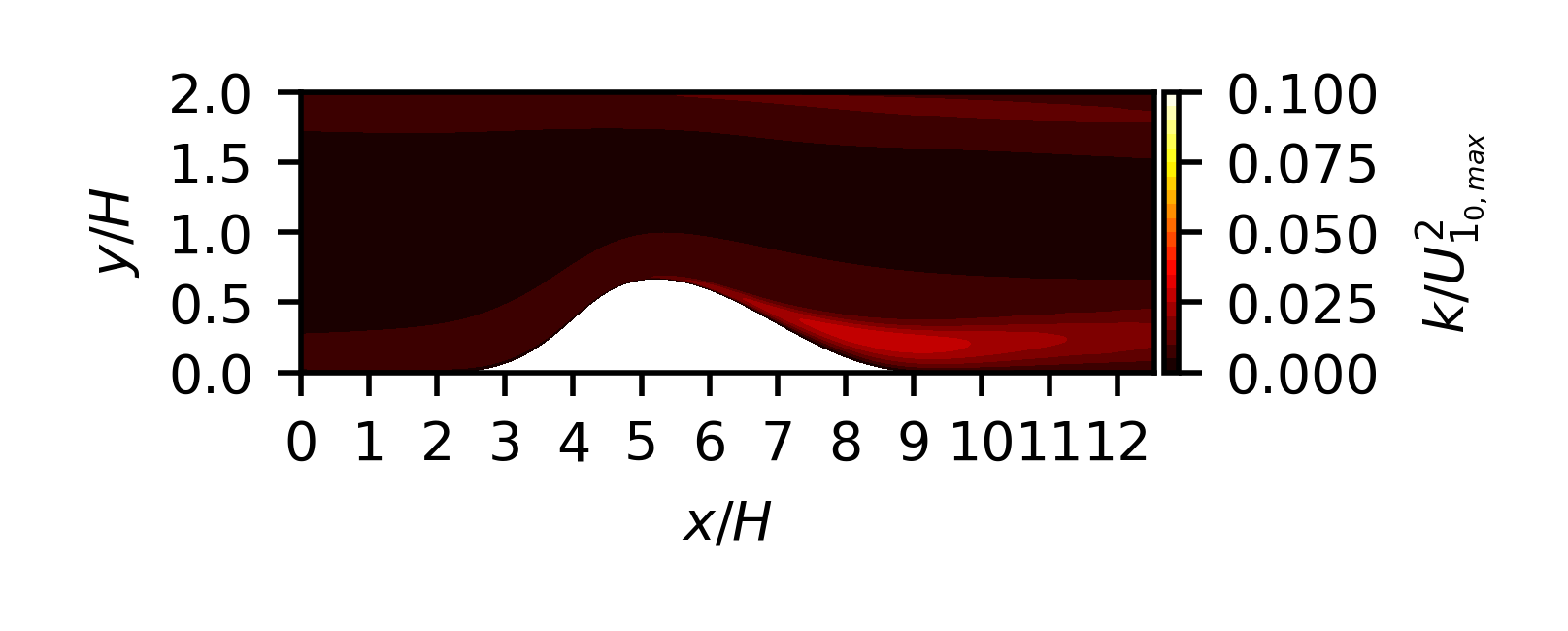}}
\end{subfloat}
\begin{subfloat}[Aiming for 3C without eigenvector rotation (\#5)]
    {\includegraphics[width=0.495\textwidth]{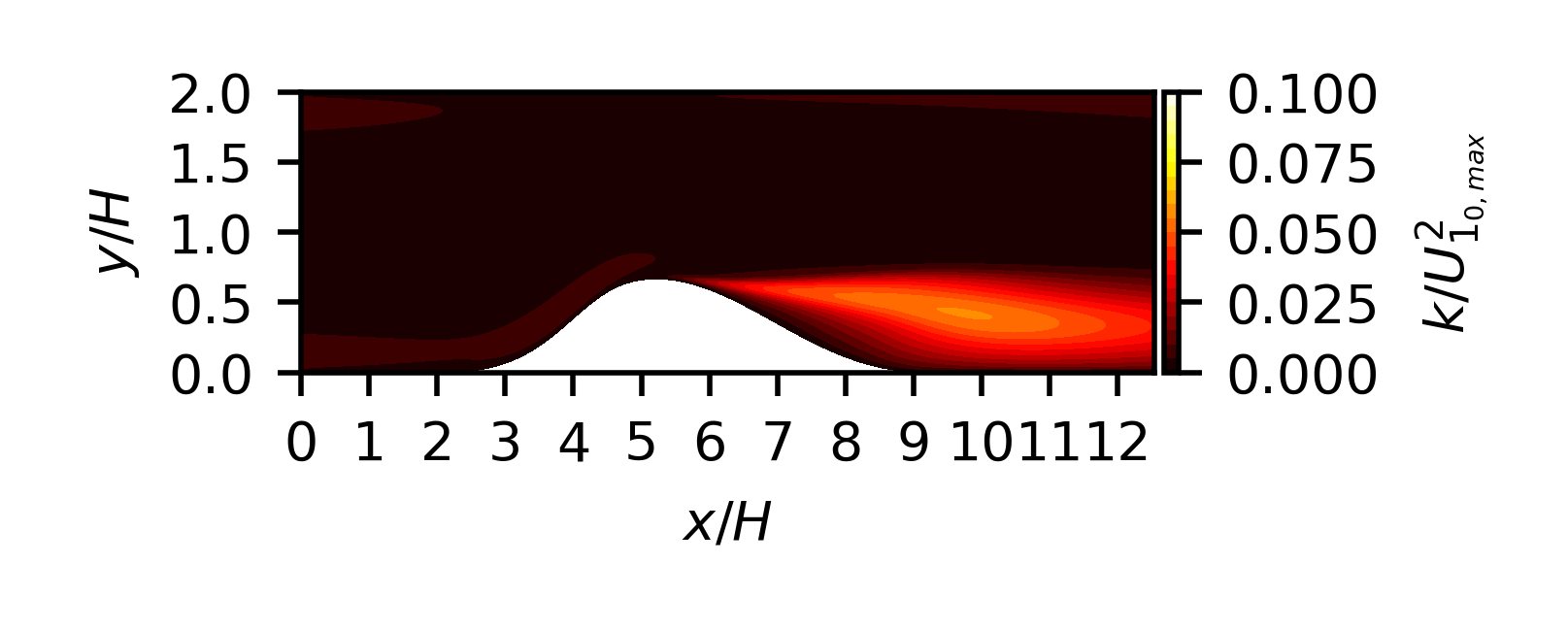}}
\end{subfloat}
\begin{subfloat}[Aiming for 3C with eigenvector rotation (\#6)]
    {\includegraphics[width=0.495\textwidth]{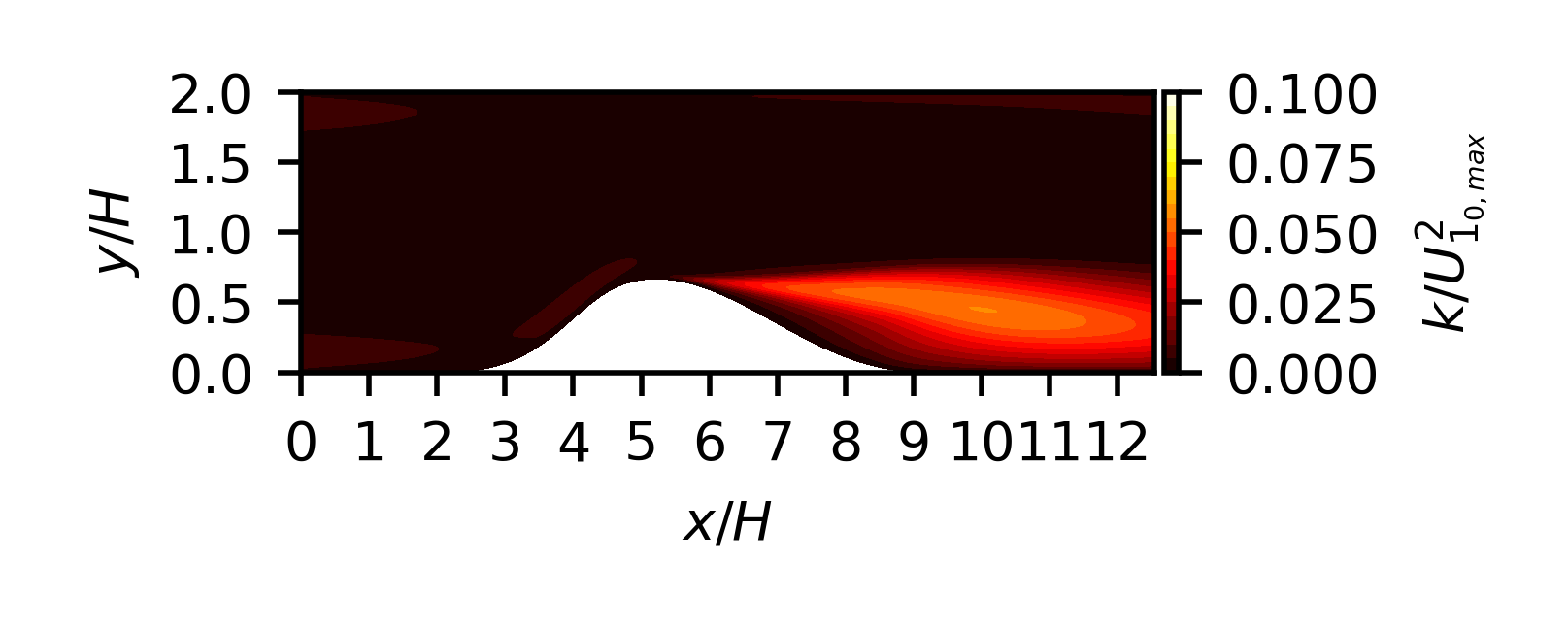}}
\end{subfloat}
\vspace{-1cm}
\caption{Evolution and comparison of turbulent kinetic energy within the converging-diverging channel between \gls{DNS}\cite{Laval} data (a), \gls{RANS} baseline (b) and \gls{EPF} simulations, applying perturbed Reynolds stress tensors (c)-(h) (see \cref{tab:conv-div_simulations}).}
\label{fig:tke_conv_div}
\end{figure}

In the subsequent section, we discuss the resulting estimated uncertainty intervals based on the eigenspace perturbation. The analysis refers to the presented \gls{QoI} in \cref{fig:uq_velocityX_conv_div} to \cref{fig:tke_conv_div}.
The estimated uncertainty for the streamwise velocity field is shown in \cref{fig:uq_velocityX_conv_div}. Perturbing the eigenspace of the Reynolds stress tensor has minor effect upstream of the diverging section ($x/H \approx 5$), where the baseline \gls{RANS} simulation closely aligns with the \gls{DNS} data. 
Due to the increased turbulent production at the one- and two-component limiting state of turbulence (as can be observed in the turbulent kinetic energy distributions in \cref{fig:tke_conv_div}), the velocity profiles become sharper with an increased gradient at the wall. 
This is also reflected in higher friction coefficients in \cref{fig:surface_cf_conv_div} at the bottom and top wall compared with the baseline simulation.
Conversely, the simulations featuring more isotropic Reynolds stresses (simulation \#5 and \#6 with $\Delta_B = 0.2$), lead to rounder velocity profiles and reduced friction coefficients.
Larger deviations between \gls{RANS} and \gls{DNS} results arise, when the flow experiences the adverse pressure gradient in the diffusor section. 
This is further reflected in an increased sensitivity of the velocity field with respect to the shape and orientation of the Reynolds stress tensor.
Due to the indirectly manipulated turbulent production behaviour, the turbulent kinetic energy evolves differently in the simulation domain for every perturbation (see \cref{fig:tke_conv_div}).
This is also in accordance with the described dependency of the turbulent production term on the eigenvalues in the front section of the diffusor (see also \cref{sec:eigenvaluePerturbation}). 
The turbulent kinetic energy production significantly affects the size of the separation bubble due to the adverse pressure gradient.
While the simulations aiming for the one- and two-component turbulence state completely suppress the separation zone at the lower wall, the simulations \#5 and \#6, featuring more isotropic turbulence both separate early and over-predict the reattachment length (see \cref{fig:surface_cf_conv_div}).
While the static wall pressure reduction in the converging section is not affected by the eigenspace perturbation, the pressure recovery in the diffusor section shows growing turbulence model uncertainty (see \cref{fig:surface_cp_conv_div}).
These uncertainty intervals on the pressure coefficient, underline the increased model-form uncertainty when it comes to adverse pressure gradient flows. 
The reduced turbulent production of the simulations \#5 and \#6 in the converging section $x\leq 5$ results in increased turbulent kinetic energy in the massively separated section (see \cref{fig:tke_conv_div}).
As already described in the theoretical parts of this paper (see \cref{sec:eigenvectorPerturbation} and \cref{sec:realizablity}), the rotation of the eigenvector matrix mainly affects the turbulent production process indirectly. 
This is especially highlighted by decreased turbulent kinetic energy patterns in the upstream section of the domain.

It is noticeable, that the simulations aiming for the two-component limiting state of the Reynolds stress tensor, show the best agreement with the \gls{DNS} data in the diffusor section for the considered \gls{QoI}.
Additionally, the \gls{DNS} data are included in the turbulence model uncertainty estimates in most of the plots, which also validates the \gls{EPF} to a certain extent, although the authors are aware of the fact, that this is not the main goal of the perturbation methodology. The interested reader is referred to the discussion on the restrictions and capabilities of the framework in Matha et al.~\cite{MathaCF}.
Regarding the potentially large uncertainty intervals concerning the considered \gls{QoI}, we must note that the eigenspace perturbations of the Reynolds stress tensor were chosen to be sufficiently large, allowing the \gls{CFD} solver to handle it just well enough to produce a convergent solution.
On the one hand, this enables exploring the capabilities of the considered \gls{EPF}, and on the other hand, it represents the most conservative estimate of turbulence model uncertainty.
\enlargethispage{0.4cm}
For upcoming design decisions considering this framework, design engineers would likely aim for a non-overly conservative estimate of turbulence model uncertainty, as they may introduce expert knowledge into the analysis.

\section{\label{sec:conclusion}Conclusion \& Outlook}
Uncertainty estimation in the context of \gls{RANS} turbulence modeling is crucial in industrial applications as it provides a quantifiable measure of the reliability of \gls{CFD} simulations.
Accurate assessment of physically plausible uncertainties ensures the credibility of simulation results, allowing engineers and designers to make informed decisions under uncertainty\cancelText{, using approaches like Design Under Uncertainty}.
The Eigenspace Perturbation Framework has established itself as the only physics-based uncertainty quantification approach for turbulence model uncertainty. It has been applied to problems in Aerospace, Civil, Environmental, Mechanical Engineering. It is widely used in leading CFD software. This underscores the need to ensure Verification \& Validation of this framework. However due to its newness, in depth verification of the rationale and application of this framework have not been conducted. This need is addressed by our investigation. \\
In this work, our primary focus centers on the eigenvector perturbation of the Reynolds stress tensor that has received limited attention in the literature.
We systematically analyze that the eigenvector perturbation may violate Reynolds stress tensor dynamics under specific conditions.
The present study derives and introduces physics-based constraints for eigenvector perturbations, adhering to the realizability and stability of the uncertainty estimation procedure. 
The application of these constraints to the flow within a converging-diverging channel illustrates improved stability and accuracy in capturing the turbulent behavior. 
The flow characteristics of this case encompass turbulent boundary layer, separation and reattachment regions, revealing deviations of \gls{RANS} and \gls{DNS} results.
Applying the \gls{EPF} unveils significant sensitivity of the considered \gls{QoI} based on Reynolds stress anisotropy and its eigenvector alignment with the strain-rate tensor.
Based on the present paper and our previous research~\cite{Matha2023}, we have successfully identified challenges in the application and interpretability and proposed potential solutions.
Our future investigations will focus on implications of the physics-constrained Reynolds stress tensor perturbation method for more complex engineering flows, such as those encountered in turbomachinery components.
This will provide valuable insights into the method's practical utility.

\begin{acknowledgments}
This paper is based on a project funded by \newTextTwo{MTU Aero Engines AG and }the German Federal Ministry for Economic Affairs and Climate Action under the funding code 03EE5041A.
The authors are responsible for the content of this  publication.
\end{acknowledgments}

\section*{Data Availability Statement}
The data that support the findings of this study are available from the corresponding author upon reasonable request.

\appendix

\section{\label{app:permutingEigenvectors} Rotation properties of eigenvector permutation~\cite{iaccarino2017eigenspace}}
In this section, we show, that the eigenvector permutation (first and third eigenvector) is identical to rotation around second eigenvector with $\pi/2$.
Any rotation around an arbitrary vector $\mathbf{v} = \left(v_1, v_2, v_3\right)^T$ by $\alpha$ can be described via the rotation matrix
\begin{equation}
\label{eq:arbitraryRotationMatrix}
\mathbf{R}= 
\begin{pmatrix}
\cos(\alpha)+v_1^2(1-\cos(\alpha)) &  v_1 v_2 (1-\cos(\alpha))-v_3 \sin(\alpha) & v_1 v_3 (1-\cos(\alpha))+v_2 \sin(\alpha)\\
v_1 v_2 (1-\cos(\alpha))+v_3 \sin(\alpha) & \cos(\alpha)+v_2^2 (1-\cos(\alpha)) & v_2 v_3 (1-\cos(\alpha))-v_1 \sin(\alpha)\\
v_1 v_3 (1-\cos(\alpha))-v_2 \sin(\alpha) &v_2 v_3 (1-\cos(\alpha))+v_1 \sin(\alpha) & \cos(\alpha)+v_3^2 (1-\cos(\alpha)) 
\end{pmatrix} \ \text{.}
\end{equation}

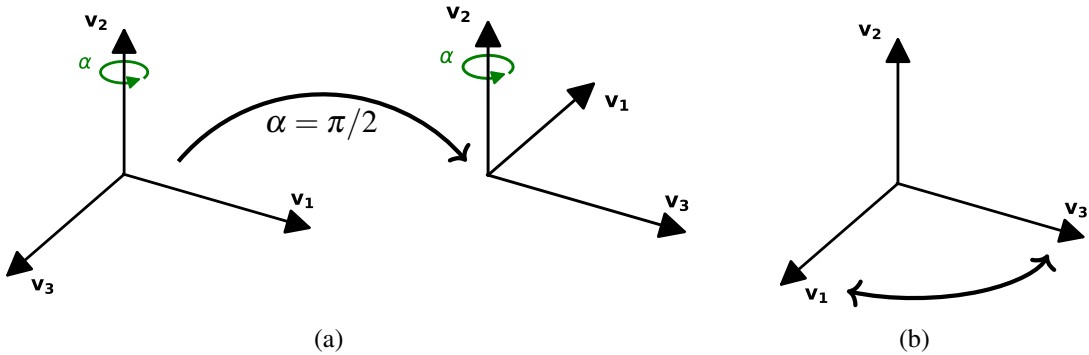
\begin{figure}[htb]
\captionsetup[subfloat]{captionskip=-10pt}
\centering
    \begin{subfloat}[]
        {\begin{tikzpicture}
            \node[] at (-3,0) {
                \includegraphics[width=0.3\textwidth, trim=2.1cm 7.5cm 1cm 7cm, clip=True]{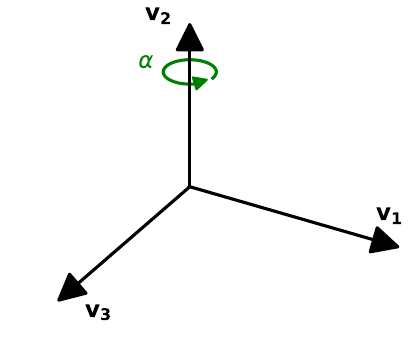}};
            \node[] at (2,0) {
                \includegraphics[width=0.3\textwidth, trim=2.5cm 7.5cm 1cm 7cm, clip=True]{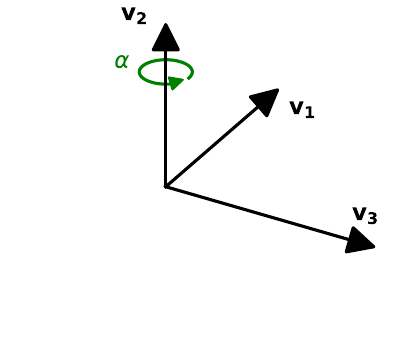}};
            \draw[->, ultra thick] (-2.5,0) 
                arc [
                    start angle=140,
                    end angle=40,
                    x radius=2.5cm,
                    y radius =2.5cm
                    ];
            \node[black] at (-0.6, 0.5) {\normalsize$\alpha = \pi/2$};
        \end{tikzpicture}}  
    \end{subfloat}
    \begin{subfloat}[]
        {\begin{tikzpicture}
            \node[] at (0,0) {
                \includegraphics[width=0.3\textwidth, trim=2.1cm 7.5cm 1cm 7cm, clip=True]{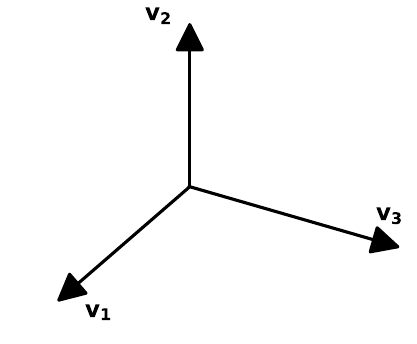}};
                \draw[<->, ultra thick] (-0.9,-1.6) 
                arc [
                    start angle=-120,
                    end angle=-10,
                    x radius=1.8cm,
                    y radius =0.7cm
                    ];
        \end{tikzpicture}} 
    \end{subfloat}
\caption{Comparison of modifying eigenvectors of the Reynolds stress tensor. (a) Rotating eigenvector matrix around second eigenvector $\mathbf{v_2}$ by $\alpha=\pi/4$. (b) Schematic representation of eigenvector perturbation by permuting first and third eigenvector~\cite{iaccarino2017eigenspace}}
\label{fig:eigenvectorRotationVSPermutation}
\end{figure}

The eigenvector matrix contains column-wise orthogonal, normalized eigenvectors $\mathbf{v_1}, \mathbf{v_2}, \mathbf{v_3}$
\begin{equation}
\mathbf{v}= 
\begin{pmatrix}
v_{11} & v_{21} & v_{31}\\
v_{12} & v_{22} & v_{32}\\
v_{13} & v_{23} & v_{33} 
\end{pmatrix} \ \text{.}
\end{equation}
The eigenvectors are ordered with respect to the respective eigenvalues in descending order.
Rotating $\mathbf{v}$ around $\mathbf{v_2}$ with $\alpha = \pi/2$ leads to:
\begin{equation}
\label{eq:rotationEqualPermuting}
\begin{split}
\mathbf{v^*} &= \mathbf{R} \mathbf{v}\\
&=
\begin{pmatrix}
v_{21}^2 &  v_{21} v_{22}-v_{23} & v_{21} v_{23}+v_{22}\\
v_{21} v_{22} +v_{23} & v_{22}^2  & v_{22} v_{23} -v_{21} \\
v_{21} v_{23} -v_{22} &v_{22} v_{23} +v_{21} & v_{23}^2  
\end{pmatrix}
\begin{pmatrix}
v_{11} & v_{21} & v_{31}\\
v_{12} & v_{22} & v_{32}\\
v_{13} & v_{23} & v_{33} 
\end{pmatrix} \\
&=\begin{pmatrix}
\left(v_{21}\left(\mathbf{v_2} \cdot \mathbf{v_1}\right)-\left(\mathbf{v_1} \times \mathbf{v_2}\right)_1\right)
& \left(v_{21} |v_2|^2\right)
& \left(v_{21}\left(\mathbf{v_2} \cdot \mathbf{v_3}\right)+\left(\mathbf{v_2} \times \mathbf{v_3}\right)_1\right)\\
\left(v_{22}\left(\mathbf{v_2} \cdot \mathbf{v_1}\right)+\left(\mathbf{v_1} \times \mathbf{v_2}\right)_2\right)
& \left(v_{22}|v_2|^2\right)
& \left(v_{22}\left(\mathbf{v_2} \cdot \mathbf{v_3}\right)-\left(\mathbf{v_2} \times \mathbf{v_3}\right)_2\right)\\
\left(v_{23}\left(\mathbf{v_2} \cdot \mathbf{v_1}\right)-\left(\mathbf{v_1} \times \mathbf{v_2}\right)_3\right)
& \left( v_{23}|v_2|^2\right)
&\left(v_{23}\left(\mathbf{v_2} \cdot \mathbf{v_3}\right)+\left(\mathbf{v_2} \times \mathbf{v_3}\right)_3\right)
\end{pmatrix}\\
&=\begin{pmatrix}
-v_{31}& v_{21}& v_{11}\\
-v_{32}& v_{22}& v_{12}\\
-v_{33}&  v_{23}& v_{13}
\end{pmatrix}
\end{split}
\end{equation}
As it can be seen in \cref{eq:rotationEqualPermuting}, the rotation result in permuting $\mathbf{v_{1}^*} = -\mathbf{v_3}$ and $\mathbf{v_{3}^*} = \mathbf{v_1}$. 
This is illustrated in \cref{fig:eigenvectorRotationVSPermutation} and compared with the formerly proposed eigenvector permutation.\
Note: In contrast to rotating the eigenvectors, simple permutation of $\mathbf{v_1}$ and $\mathbf{v_3}$ evidently lead to a left-handed oriented eigenvector matrix. 
The resulting Reynolds stress tensor, when reconstructed based on \cref{eq:spectralDecompositionR*}, is identical due to the characteristic of the spectral decomposition.

\clearpage

\section{\label{app:convDivSketch} Schematics of converging-diverging flow example}
\begin{figure}[htb]
\captionsetup[subfloat]{captionskip=-5pt}
\centering
\begin{subfloat}[]
    {\includegraphics[width=\textwidth]{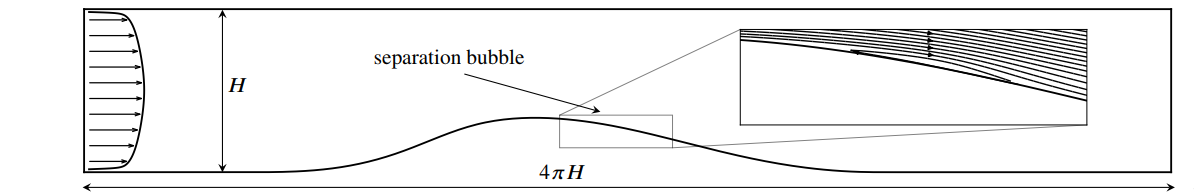}}
\end{subfloat}
\begin{subfloat}[]  
    {\includegraphics[width=\textwidth]{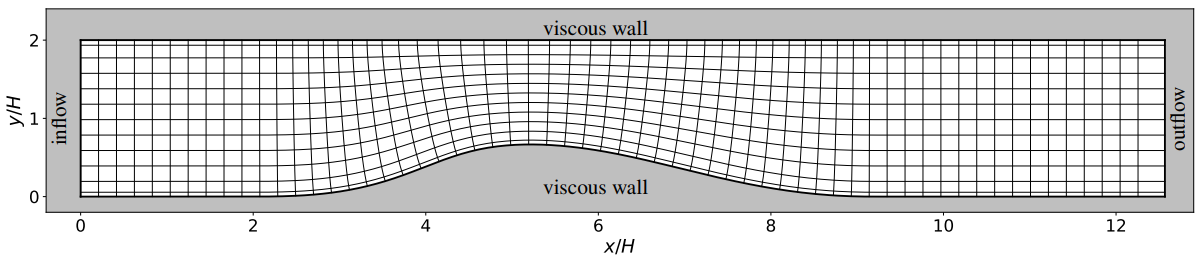}}
\end{subfloat}
\begin{subfloat}[]  
    {\includegraphics[width=0.97\textwidth]{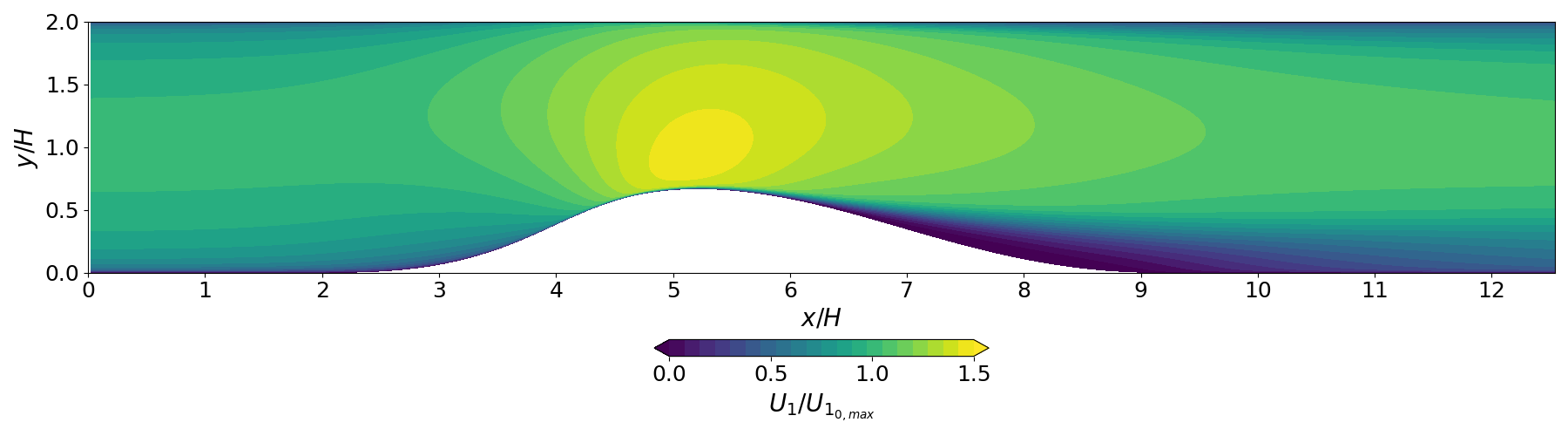}}
\end{subfloat}
\caption{\label{fig:convDivSketch}\newTextTwo{Introduction of}\cancelTextTwo{Schematic} converging-diverging setup. (a) Relative dimensions and sketch of the flow \gls{DNS} data~\cite{Laval} based. (b) Mesh \newTextTwo{consisting of 242x242x1 grid points} (every fourth line in streamwise direction and every twentieth line in wall normal direction shown) and boundary conditions; slip conditions/inviscid walls are applied in spanwise direction. \newTextTwo{(c) Streamwise velocity based on the \gls{RANS} baseline computation using Menter SST and the mesh presented in (b).}}
\label{fig:convDiv}
\end{figure}

\section{\label{app:failingPerturbation} Instability introduced by former eigenvector permutation}

\begin{figure}[tb]
 \centering
\includegraphics[width=\textwidth]{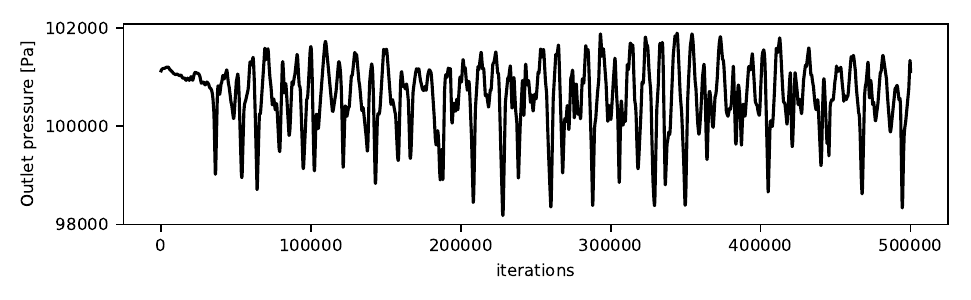}%
\vspace{-1cm}
\caption{\label{fig:fail:pressure}Evolution of the area averaged outlet pressure over iteration count for the simulation using eigenvector permutation without any eigenvalue perturbation.}
\end{figure}

\begin{figure}[htb]
 \centering
\includegraphics[width=\textwidth]{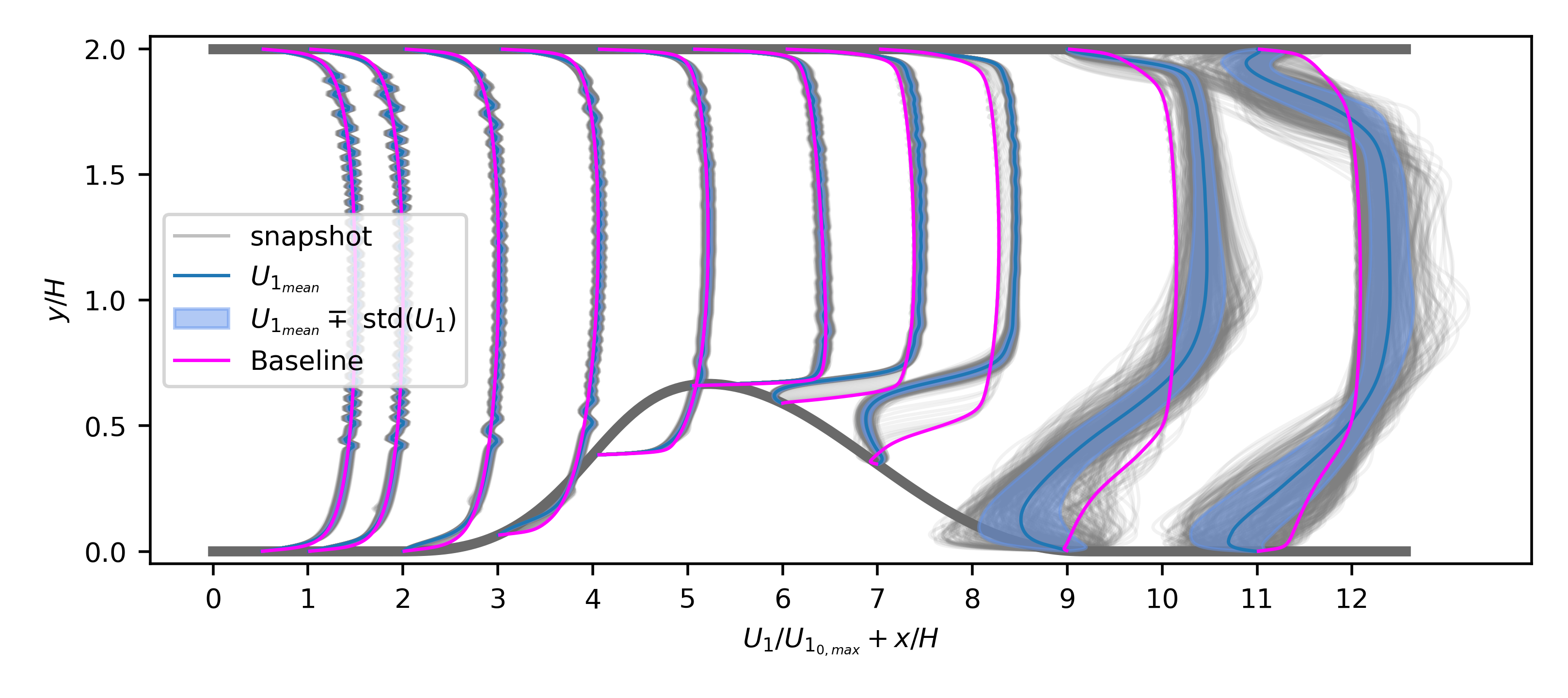}%
\vspace{-1cm}
\caption{\label{fig:fail:velocity} Streamwise velocity inside of the converging-diverging channel based on pure eigenvector permutation without any eigenvalue perturbation. The snapshots are taken every 1000 iteration, while the mean $\overline{U_1}$ and the standard deviation std($U_1$) are determined between iterations 400.000 to 500.000.}
\end{figure}

We apply the eigenvector permutation \cite{iaccarino2017eigenspace} in the former implementation of the \gls{EPF} without any eigenvalue perturbation ($\Delta_B =0$).
When checking the evolution of the outlet pressure in \cref{fig:fail:pressure} (specified mass flow rate is specified as outlet boundary condition), it becomes evident, that the simulation is unstable. 
As a consequence, the streamwise velocity reveals significant variations variations at each snapshot in \cref{fig:fail:velocity}. Additionally, the application of non-realizable Reynolds stress tensor dynamics, creates non-physical countergradient transport (see \cref{sec:realizablity}) , which results in the zigzag like velocity profiles.

\begin{figure}[htb]
 \centering
\includegraphics[width=\textwidth]{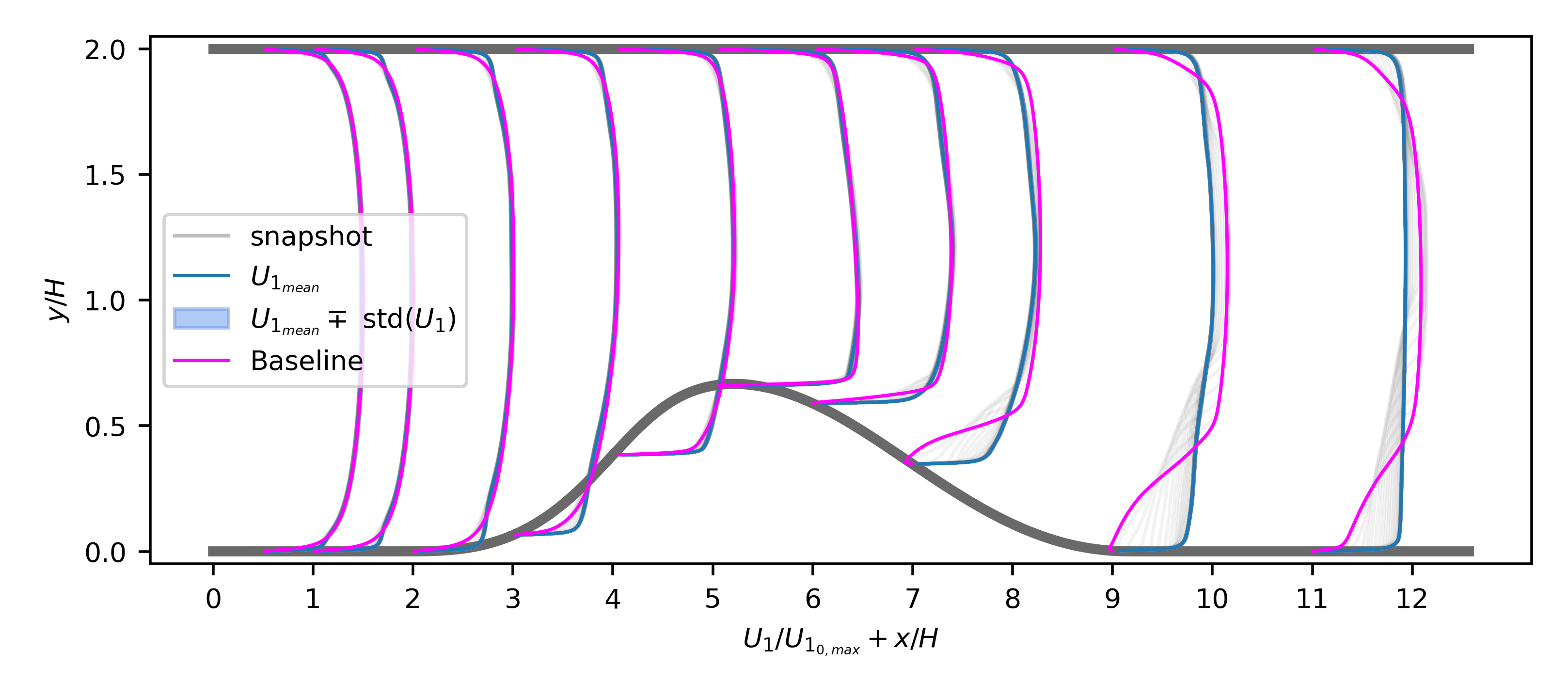}%
\vspace{-1cm}
\caption{\label{fig:works:velocity} Streamwise velocity inside of the converging-diverging channel for simulation \#2 (see \cref{tab:conv-div_simulations}) using moderated eigenvector perturbation and eigenvalue modification towards the one-component turbulent limiting state. The snapshots are taken every 1000 iteration, while the mean $\overline{U_1}$ and the standard deviation std($U_1$) are determined between iterations 400.000 to 500.000.}
\end{figure}

In contrast, the streamwise velocity snapshots of the perturbed simulations, used in \cref{sec:application}, converge over runtime of the simulations (\cref{fig:works:velocity} presents one example).

\clearpage

\bibliography{aipsamp}
\end{document}